\documentclass[lettersize,journal]{IEEEtran}
\usepackage{amsmath,amsfonts}
\usepackage{algorithmic}
\usepackage{algorithm}
\usepackage{array}
\usepackage[caption=false,font=normalsize,labelfont=sf,textfont=sf]{subfig}
\usepackage{booktabs}
\usepackage[most]{tcolorbox}
\usepackage{multirow}
\usepackage{textcomp}
\usepackage{stfloats}
\usepackage{url}
\usepackage{verbatim}
\usepackage{graphicx}
\usepackage{cite}
\usepackage{csvsimple}
\usepackage{tabularx, makecell}
\usepackage{times}
\usepackage{latexsym}
\usepackage{graphicx}   
\usepackage{xspace}   
\usepackage{multirow}
\usepackage{float}
\usepackage{array}
\usepackage{tabularx}
\usepackage{mdframed}
\usepackage{xcolor}
\usepackage{placeins}  
\usepackage[most]{tcolorbox}
\usepackage{booktabs}
\usepackage{enumitem}
\usepackage{caption}
\usepackage{subcaption}
\usepackage{amssymb}
\usepackage{pifont}
\usepackage{textcomp} 
\usepackage{listings}
\usepackage{csvsimple}
\usepackage{afterpage}  
\usepackage{url} 
\usepackage{adjustbox}
\usepackage{lipsum,graphicx,multicol}
\usepackage[table]{xcolor}
\usepackage{cite}
\usepackage{changepage} 
\usepackage{multirow}
\usepackage{makecell}  
\usepackage{booktabs}
\usepackage[hidelinks]{hyperref}

\begin{document}


\title{\textbf{\Large Can LLM Prompting Serve as a Proxy for Static Analysis in Vulnerability Detection} \\[1ex] \large Partial Code Analysis with LLM-Based CWE-Specific Rules: Addressing a Critical Security Challenge in an AI-Driven Era}

\newcommand{\AuthorBlock}[4]{%
  \begin{tabular}[t]{@{}c@{}}
    #1\\                                   
    \textit{#2}\\                          
    #3\\                                   
    #4                                     
  \end{tabular}}

\author{%
\begin{tabular*}{\textwidth}{@{\extracolsep{\fill}}ccc}  
  \AuthorBlock{Ira Ceka}{Columbia University}{New York, New York}{iceka@cs.columbia.edu} &
  \AuthorBlock{Feitong Qiao \textsuperscript{*} \thanks{ \textsuperscript{*} Equal contribution}}{Columbia University}{New York, NY, USA}{flq2101@columbia.edu} &
  \AuthorBlock{Anik Dey \textsuperscript{*}}{Amherst College}{Amherst, MA, USA}{andey25@amherst.edu} \\[4.5em]
  
  \AuthorBlock{Aastha Valecha}{Columbia University}{New York, NY, USA}{av3180@columbia.edu} &
  \AuthorBlock{Gail Kaiser}{Columbia University}{New York, NY, USA}{kaiser@cs.columbia.edu} &
  \AuthorBlock{Baishakhi Ray}{Columbia University}{New York, NY, USA}{rayb@cs.columbia.edu}
\end{tabular*}
}

\maketitle

\begin{abstract}
Despite their remarkable success, large language models (LLMs) have shown limited ability on safety-critical code tasks such as vulnerability detection. Typically, static analysis (SA) tools, like CodeQL, CodeGuru Security, etc., are used for vulnerability detection. SA relies on predefined, manually-crafted rules for flagging various vulnerabilities. Thus, effectiveness of SA in detecting vulnerabilities depends on human experts and is known to report high error rates. In this study we investigate whether LLM prompting can be an effective alternative to these static analyzers in the partial code setting. We propose prompting strategies that integrate natural language instructions of vulnerabilities with contrastive chain-of-thought reasoning, augmented using contrastive samples from a synthetic dataset. Our findings demonstrate that security-aware prompting techniques can be effective alternatives to the laborious, hand-crafted rules of static analyzers, which often result in high false negative rates in the partial code setting. When leveraging SOTA reasoning models such as DeepSeek-R1, each of our prompting strategies exceeds the static analyzer baseline, with the best strategies improving accuracy by as much as 31.6\%, F1-scores by 71.7\%, pairwise accuracies by 60.4\%, and reducing FNR by as much as 37.6\%.
\end{abstract}


\section{Introduction}
\label{sec:intro}
The rise of large language models (LLMs) has led to dramatic improvements in natural language processing (NLP), demonstrating remarkable performance across various tasks, such as language understanding, translation, text generation, and text summarization. These advancements have extended into the domain of code, facilitating complex tasks such as code generation, code summarization, and code repair~\cite{touvron2023llama, li2023starcoder}.

One particularly critical yet challenging task in the code domain is \textit{vulnerability detection}, which is an ever-growing concern 
by software security practitioners and users. The task of vulnerability detection can be thought of a binary classification task---given a piece of code, a detector will predict vulnerable/non-vulnerable. 

Traditionally, researchers used a range of program analysis techniques, including static and dynamic analysis~\cite{chakraborty2021deep}.
Static analysis (SA) is a method of analyzing code without executing it; SA relies on a set of predefined rules, or specifications. 
These specifications define code in terms of \textit{sources} (functions that provide unsafe inputs), \textit{sinks} (points in the code where unsafe data is used), and \textit{sanitizers} (mechanisms for cleaning and validating the input to make it safe for use).
Developers and security engineers must \textbf{manually craft} and maintain these specifications, often relying on their domain expertise and existing documentation. This laborious process can often result in overlooked rules and incomplete analysis.
It is well-known that static analysis tools are ~\textbf{error-prone}~\cite{Kang_2022}. Hand-crafted rules can lead to over-approximation or flagging of context that is wrongly-assumed to be exploitable. This often results in high false positive rates. In practice, false alarms incur significant costs by diverting engineering resources to investigating findings that are later determined to not be vulnerabilities after review.

In addition to these challenges, static analysis tools struggle with partial code. Yet, detecting vulnerabilities in partial code is an increasingly relevant practical setting. Throughout code development, developers routinely reuse functions or snippets of code from open-source solutions ~\cite{yang2017stackoverflowgithubsnippets, Jahanshahi_2025, chen2024developers}. During pull-request reviews, code changes are presented in fragments, requiring developers to frequently review and validate partial code. In the domain of secure code generation, partial code must be evaluated for safeguarding as it is being written. Malicious actors cannot submit full repositories to AI chat-bots or LLM-based coding assistants; instead, they provide partial code and expect partial code back. Consequently, detecting malicious input and preventing vulnerable code from being generated are critical tasks ~\cite{sahai2025amazon}. Furthermore, recent industry reports indicate that roughly 30\% of code being written at large companies is now generated by AI ~\cite{msn2025ai_code}. Recent studies have found vulnerability risks in AI-generated code ~\cite{cotroneo2024vulnerabilities, hamer2024just, khoury2023secure}, with some indicating that up to 40\% of code produced by certain coding assistants contains security flaws ~\cite{pearce2025asleepatkeyboardu}. With this, there is a significant risk that insecure code propagates into production systems, if left undetected ~\cite{sahai2025amazon}.

To overcome the limitations of static analysis, many research works have tried to apply Language Model-based approaches~\cite{zhou2019devign, chakraborty2021deep, li2021vulnerability, hin2022linevd}. However, recent studies found that LM-based approaches (both zero and few-shot settings) exhibit limited performance in zero-shot settings \cite{steenhoek2024comprehensive, ding2024vulnerability}. While some ML-based techniques show promise on synthetic datasets, their performance remains limited on high-quality, real-world data \cite{ullah2024llms, khare2023understanding}. This gap underscores the ongoing challenge of leveraging LLMs to accurately identify security vulnerabilities.

However, existing work has shown that fine-tuning and training vulnerability detection models is computationally expensive ~\cite{shestov2025finetuning,lu2024grace,li2021vulnerability}, and their performance is further constrained by the limited availability of high-quality training datasets ~\cite{lu2024grace,li2021vulnerability}. In contrast, we present an approach that can elicit the localities of the patterns learned in training within LLMs in a self-generated manner. 
The main problem with these LLM-based techniques are they only rely on data-driven approaches, ignoring all the domain-specific properties associated with different types of vulnerabilities. 

To address these limitations, we develop prompting strategies that incorporate natural language descriptions of vulnerabilities, allowing the model to understand and reason about potential attack surfaces and threat behaviors. By adopting a contrastive chain-of-thought approach, the model evaluates multiple lines of reasoning in parallel to differentiate between correct and erroneous vulnerability assessments. We reinforce these reasoning steps using contrastive samples from a synthetic dataset, providing clear examples of both valid and invalid vulnerability instances. This combined method aims to improve LLM's  capability in identifying security flaws across diverse codebases.

Our findings indicate that such security property-aware prompting strategies significantly outperform state-of-the-art static analysis tools such as CodeGuru Security\footnote{\url{https://aws.amazon.com/codeguru/features/}} in the partial code setting. We demonstrate the effectiveness of these techniques through rigorous evaluation, showing promising results in improving over static analysis tools. 

In particular, we evaluated using several vulnerability detection datasets; these composed of both in-the-wild SVEN ~\cite{he2023large}, CVEFixes (diverse programming languages), CVEFixes C/C++ ~\cite{bhandari2021cvefixes} and synthetic datasets: Juliet C/C++ ~\cite{vishnyakov21}. We investigate on several large language models: namely, {o1}, {Claude 3.5 Sonnet v2}, and {DeepSeek-R1}.  
Experimental evaluation shows that our prompting techniques perform well against both the prompting and static analyzer baselines across different evaluation metrics. When using state-of-the-art reasoning models like DeepSeek-R1, each prompting strategy outperforms the static analyzer baseline. Our most effective strategies boost accuracy of vulnerability detection by up to 31.6\%, F1-scores by 71.7\%, pairwise accuracies by 60.4\%, and reduce FNR by up to 37.6\%.
Thus, we show that our domain-specific prompting strategies can be used as a complementary approach to detect vulnerabilities.

In summary, this paper makes the following contributions:
\begin{itemize}
 \setlength{\itemsep}{0pt}
 \setlength{\parskip}{0pt}
\item We demonstrate the effectiveness of prompting strategies over static analysis rules for vulnerability detection in the partial code setting

\item We evaluate on a full high-quality, clean, reliable dataset such as SVEN. 
\item We show improvements over critical metrics such as pairwise accuracy, aiming to increase detection rates while reducing false alarms. 
\item We propose a method for utilizing CWE-specific open-source information to generate precise natural language descriptions of various vulnerability types. \footnote{These generated descriptions are then validated by cybersecurity experts with specialized domain knowledge in CWE vulnerability types, ensuring their accuracy and reliability.}
\item We compare LLM performance to that of static analyzers like CodeGuru Security and Semgrep. 
\item We provide a comprehensive analysis of the strengths and limitations of LLMs in identifying vulnerabilities through a detailed examination of over 414 LLM responses.

\end{itemize}
\section{Background} We introduce and define relevant terms related to vulnerability detection.
\subsection{Vulnerability Terms}

\paragraph{Vulnerable} This label indicates a code that has a security flaw or weakness that can be exploited by an attacker. A vulnerable piece of code is at risk of being the entry point or cause of a security breach. These vulnerabilities are often categorized by their types (see \ref{sec:cwe2.2}).

\paragraph{Fixed} A term used when a previously identified vulnerability in the code has been "remedied" or fixed. The fix adjusts the "vulnerable" code to eliminate the security weakness. A fragment of code that has been "fixed" for a particular CWE is considered non-vulnerable to that CWE; thus, we may use the terms interchangeably throughout the paper.

\paragraph{Non-vulnerable} This label indicates code that has no vulnerabilities detected within the specific checks conducted. However, this does not guarantee the code is entirely secure; it only means no issues were found within the tested scope.

\subsection{CWE: Common Enumeration Weaknesses}
\label{sec:cwe2.2}
We direct our efforts on 4 critical categories for which we have data across all of our datasets. These appear in MITRE's CWE Top 25 Most Dangerous Software Weaknesses. \footnote{\url{https://cwe.mitre.org/top25/}}
\begin{itemize}
    \setlength{\itemsep}{0pt}
    \setlength{\parskip}{0pt}
    \item \textbf{CWE-78 (OS Command Injection)}: occurs when external user inputs are directly passed into system commands without proper sanitization.
    \item \textbf{CWE-190 (Integer Overflow)}: occurs when an arithmetical operation on integers results in a value outside of the variable's allowable range, leading to buffer overflows or sensitive information leakage.
    \item \textbf{CWE-476 (Null Pointer Dereference)}: occurs when an application attempts to dereference a pointer it assumes to be valid, but is actually NULL, leading to a crash or unexpected termination.
    \item \textbf{CWE-416 (Use After Free)}: occurs when an application accesses memory after it has been freed, leading to exploitable conditions.
\end{itemize}

\section{Prompting Strategies}

In this section, we illustrate our prompting strategies.  
To generate detailed vulnerability explanations with GPT-4, we use contrastive Juliet C/C++ samples and specialized prompts. We select vulnerable and fixed code pairs (Phase 1), prompt the LLM for CWE-specific detection instructions (Phase 2), and generate chain-of-thought explanations comparing the pairs (Phase 3). Finally, we synthesize a contrastive template highlighting vulnerabilities and fixes (Phase 4). See \ref{sec:how_llm} for detail.

\subsection{Vanilla Prompt}~\label{problem}

\begin{figure}[!ht]
\centering
\includegraphics[width=0.75\columnwidth]{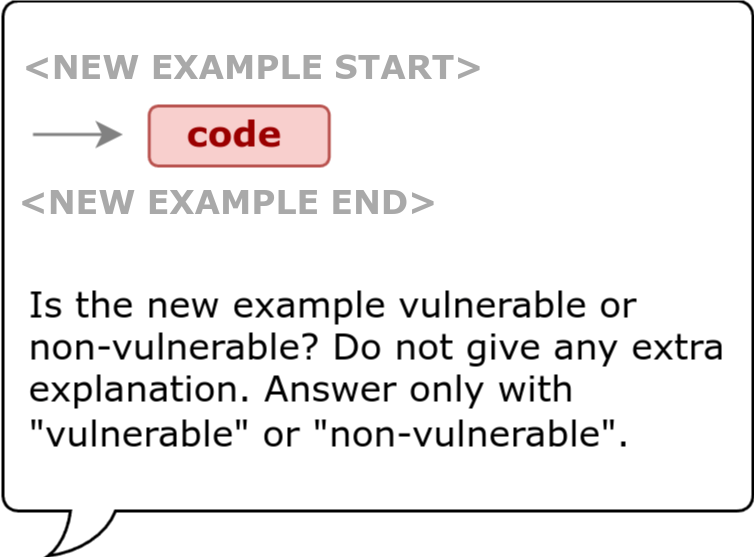}

\caption{Basic Prompt (B)}
\label{fig:basic-prompt}
\end{figure}

We design a basic “vanilla” prompt. As illustrated in Figure~\ref{fig:basic-prompt}, the code sample to be evaluated is presented before the instruction. To reduce ambiguity while ensuring that the model produces sufficient and meaningful responses, the prompt explicitly asks: “Is the example vulnerable or non-vulnerable?” In addition, the prompt includes the directive: “Do not provide any extra information…”. The inclusion of this instruction was motivated by insights from our preliminary pilot study, which revealed that without such constraints, the model frequently generated irrelevant or verbose outputs. \footnote{Our pilot study showed that without this explicit instruction, the model frequently produces superfluous and inconclusive results. Moreover, we find that tag-based formatting does not significantly impact overall results. See \ref{sec:tagbasedformatting})}

\subsection{Natural Language Instructions}~\label{sec:var} 

 In the natural language instruction setup, we add to Fig. \ref{fig:basic-prompt} by incorporating domain-specific, vulnerability-focused instructions through several distinct settings. In the NL (S0) setting, we provide the specific CWE-ID as signal. In the NL (S1) setting, the instructions are automatically generated by a large-scale language model (GPT-4). In the NL (S2) setting, the instructions are generated from examples. In the NL (S3) setting, we use real-world, human-authored instructions drawn from established security platforms such as MITRE \footnote{\url{https://cwe.mitre.org/about/index.html}}. Together, these progressive instruction styles allow us to systematically examine how prompt complexity and source of authority influence performance.

\textbf{NL (S0)}: The model is given only the CWE-ID in a zero-shot prompt. For example:~\textit{Is the new example vulnerable or non-vulnerable to CWE-190: Integer Overflow or Wraparound?}. The model must rely on its inherent training to classify the example.

\textbf{NL (S1)}: In the first setting, we utilize the LLM to generate the instructions for detecting a specific CWE, leveraging its \textit{inherent} understanding of CWEs to provide insightful and relevant instructions. 
    
\textbf{NL (S2)}: We utilize LLM-generated instructions from few-shot samples. We ask the LLM to culminate a set of instructions based on 3 pairwise (both vulnerable and fixed) samples extracted from the SVEN validation set for that particular CWE.  This method allows us to closely align the generated instructions with a practical, high-quality real-world dataset of vulnerabilities. 
    
\textbf{NL (S3)}: We opt for an approach that uses an exact match text from the instructions provided by MITRE for that particular CWE. This setting ensures that the instructions are directly correlated with authoritative, human-written, and widely-recognized instructions of vulnerabilities. These settings are not developed within a contrastive CoT framework, as we will describe in Section~\ref{sec:con}, hence they do not incorporate synthetic examples. 

\begin{figure}[htb]
\centering
\includegraphics[width=0.75\columnwidth]{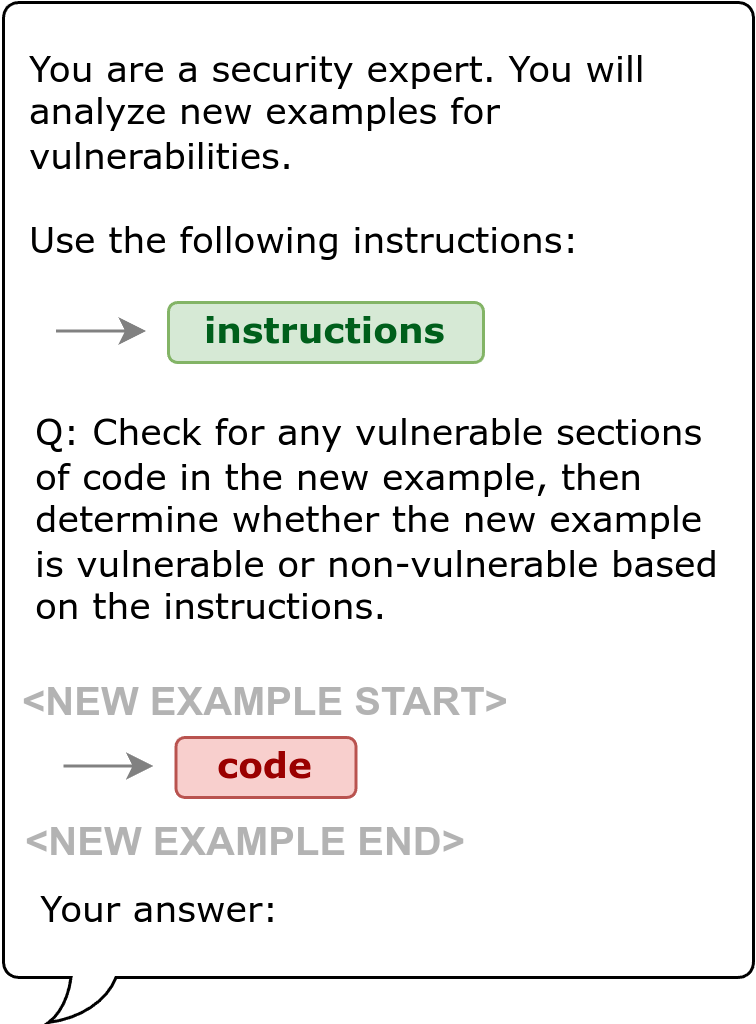}

\caption{Prompting strategy with NL instructions}
\label{fig:fig-nl}
\end{figure}

\subsection{Natural Language Instructions with Contrastive Chain of Thought}~\label{sec:con} 
\textbf{Contrastive examples.} 
We extract contrastive code pairs from the Juliet C/C++ Dynamic Test Suite. Each pair includes a vulnerable example (exhibiting a specific CWE vulnerability) and its corresponding fixed version.

\textbf{Contrastive CoT (Chain-of-thought).} 

\begin{figure}[!ht]
\centering

\includegraphics[width=0.8\columnwidth]
{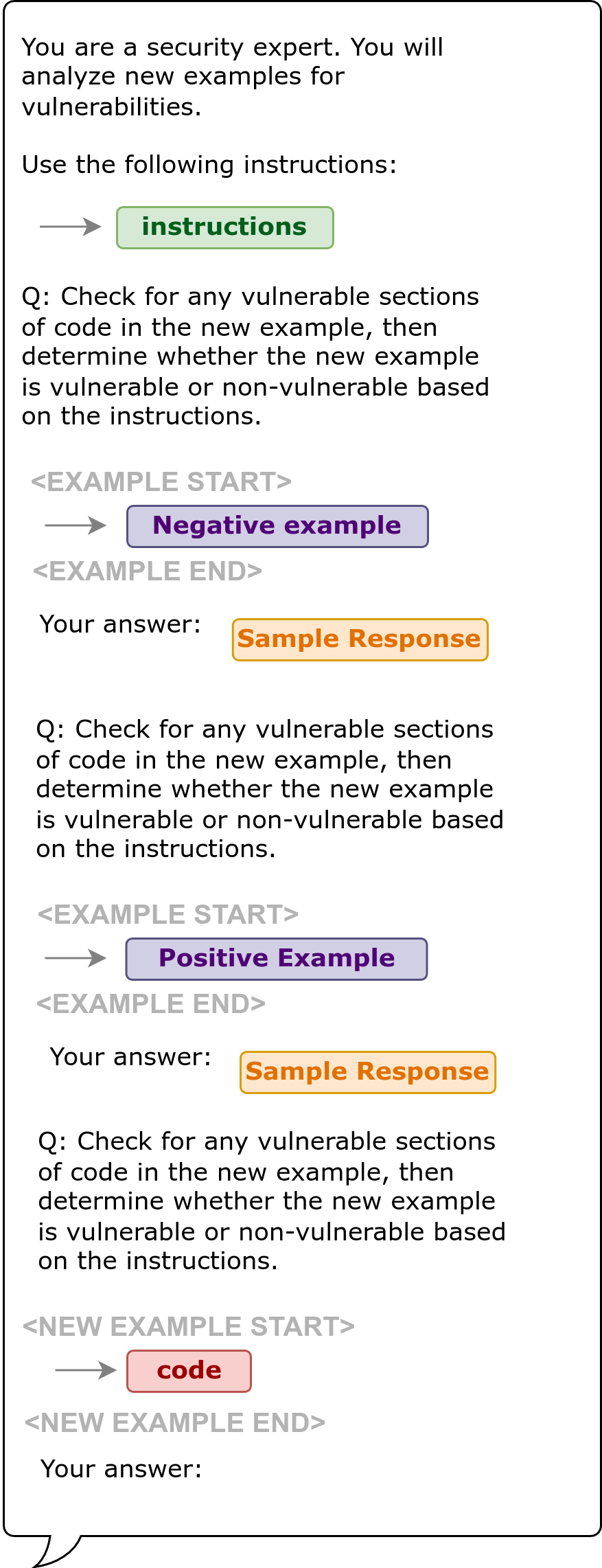}

\caption{\textbf{NL+CoT Prompt}. An example prompt provided by us during the prompt construction.}
\label{fig:fig3}
\end{figure}

 We use GPT-4 to generate the reasoning for why the synthetic example is vulnerable or fixed. Then, we embed both the synthetic example and its reasoning into a structured Q-A template. Figure~\ref{fig:fig3} shows a contrastive CoT template.
We combine NL instructions with CoT in 4 different settings:

\textbf{NL+Cot (S0)}: This is the generic chain-of-thought (CoT) setting. It includes a human-written instruction of the vulnerability from MITRE. It includes a non-contrastive CoT example and a corresponding CoT explanation. The example is the demonstrative instance from MITRE, and the explanation is the reasoning of the demonstrative example, which is written by a security professional. \footnote{Example MITRE instruction with demonstrative example: \url{https://cwe.mitre.org/data/definitions/190.html}} 

\textbf{NL+CoT (S1)}: The first setting revolves around formatting the instructions for a specific CWE as "tests" or "checks" that need to be satisfied. These instructions are crafted by the LLM, then embedded into a contrastive CoT template. 

\textbf{NL+CoT (S2)}: In the second setting, we maintain the "tests" or "checks" format but enrich the instruction generation process by feeding the official MITRE instructions alongside demonstrative examples from MITRE's comprehensive database. 

\textbf{NL+CoT (S3)}: The LLM generates instructions in free form (not as tests or checks). The instructions are embedded into a template, along with contrastive examples and LLM-generated reasoning on how the examples fit/fail the instructions. 

\begin{table*}[t] 
\centering
\caption{The categories of prompting strategies and their settings.}
\scriptsize
\resizebox{\textwidth}{!}{
\begin{tabular}{c|l|l}
\toprule
\textbf{Category} & \textbf{Setting} & \textbf{Description} \\ \midrule
Basic (B) &
  \begin{tabular}[c]{@{}l@{}}
    N/A \\[0.5em]
  \end{tabular} &
  \begin{tabular}[c]{@{}l@{}}
    Prompting: “Is the new example vulnerable or non-vulnerable?” \\
    No inclusion of vulnerability NL instruction or chain-of-thought.
  \end{tabular} \\ \midrule
\multirow{4}{*}{\begin{tabular}[c]{@{}c@{}}NL Instruction \\ describing \\ the CWE\end{tabular}} &
  Setting 0 & "Is the new example vulnerable or non-vulnerable to $<$CWE-ID$>$?" \\ \cmidrule{2-3} 
 &
  Setting 1 &
  NL instruction generated by the LLM. \\ \cmidrule{2-3} 
 &
  Setting 2 &
  NL instruction generated by the LLM after being given 3 few-shot examples from SVEN. \\ \cmidrule{2-3} 
 &
  Setting 3 &
  NL instruction provided from MITRE CWE Dictionary. \\ \midrule
\multirow{4}{*}{\begin{tabular}[c]{@{}c@{}}NL Instruction \\ + Chain-of-Thought \\ (NL+CoT)\end{tabular}} &
  Setting 0 &
  \begin{tabular}[c]{@{}l@{}}
    \textbf{Generic CoT.} NL instruction with demonstrative example and reasoning from MITRE.
  \end{tabular} \\ \cmidrule{2-3} 
 &
  Setting 1 &
  \begin{tabular}[c]{@{}l@{}}
    NL instruction generated by the LLM; formatted as “tests”/“checks”.\\
    Contrastive chain-of-thought examples provided along with LLM-generated reasoning as to how they satisfy the “checks.”
  \end{tabular} \\ \cmidrule{2-3} 
 &
  Setting 2 &
  \begin{tabular}[c]{@{}l@{}}
     NL instruction generated by the LLM given MITRE CWE Dictionary +\\ 
    MITRE demonstrative examples; formatted as “tests”/“checks”.\\
    Contrastive chain-of-thought examples provided along with LLM-generated reasoning as to how they satisfy the “checks.”
  \end{tabular} \\ \cmidrule{2-3} 
 &
  Setting 3 &
  \begin{tabular}[c]{@{}l@{}}
    NL instruction generated by the LLM; free-form format.\\
    Contrastive chain-of-thought examples provided along with LLM-generated reasoning as to how they satisfy the instruction.
  \end{tabular} \\ \bottomrule
\end{tabular}
}
\label{tab:strategies}
\end{table*}

\subsection{How We Leverage LLMs for Prompt Generation}
\label{sec:how_llm}

We illustrate the step-by-step process for how the prompts are generated in the most comprehensive case -- Natural Language + CoT. This provides clarity of insight into how we harness LLMs at each step. 

Our approach involves a multi-phase process where we leverage manually curated data from the Juliet C/C++ dataset along with LLM-generated explanations (using ChatGPT-4). The process is divided into 4 key phases:

\paragraph{\textbf{Phase 1: Pairwise Sample Selection}}

\newblock{
\begin{tcolorbox}[title=Phase 1: Input(s) and Output(s), coltitle=white, fonttitle=\bfseries]
\textbf{Input}: CWE-ID

\textbf{Output}: Synthetic vulnerable and fixed code samples from the Juliet C/C++ dataset
\end{tcolorbox}
}
In the first phase, we manually select pairwise samples for each CWE of interest from the Juliet C/C++ dataset. 

In the Juliet C/C++ Dynamic Test Suite, the synthetic pairwise samples are organized into files, with \href{https://github.com/ispras/juliet-dynamic/blob/6b876027be4345abdad78a2db156b8040c2df39a/testcases/CWE78_OS_Command_Injection/s08/CWE78_OS_Command_Injection__wchar_t_file_w32_execv_66a.c}{\textbf{each file}} representing a distinct vulnerability scenario.
These vulnerability scenarios are constructed in various ways to demonstrate or expose vulnerabilities (as well as their fixes) by combining different variations of \textbf{good sources} (safe input data), \textbf{bad sources} (insecure input or actions), and \textbf{bad sinks} (unsafe downstream commands or operations). 

This variability illustrates how different scenarios can lead to vulnerabilities and how they can be mitigated. We randomly select one such scenario for each CWE and locate the pairwise samples in the file (Ex. we have provided an example of such a pair for CWE-78: OS Command Injection in \ref{sec:juliet_samples}. Note: though a bad sink is referenced in the non-vulnerable sample, it is abstracted and reached only indirectly; the payload itself is sanitized)

\paragraph{\textbf{Phase 2: Instruction Generation with LLM}}
\newblock{
\begin{tcolorbox}[title=Phase 2: Input(s) and Output(s), coltitle=white, fonttitle=\bfseries]
\textbf{Input}: CWE-ID, a prompt

\textbf{Output}: LLM-generated instructions for vulnerability detection
\end{tcolorbox}
}

In the second phase, we prompt the LLM (ChatGPT-4) to generate a set of instructions specific to each CWE. The process begins with a predefined prompt, which includes the CWE-ID and a consistent format for instruction generation. For example, the prompt for CWE-78: OS Command Injection is structured as follows:

\begin{tcolorbox}[title=Phase 2 Prompt, colframe=blue!50!black, colback=white, colbacktitle=blue!50!black, coltext=black!80!blue, fonttitle=\bfseries] 
    You are a security expert in vulnerability detection. You specialize in identifying CWE-78: OS Command Injection. Give a set of instructions that can be used in a code review in order to identify CWE-78: OS Command Injection.
\end{tcolorbox}

We perform an additional post-processing step using an exemplar set of instructions for CWE-416, which is verified for accuracy and clarity. This exemplar serves as a template only for \textit{formatting} the instructions consistently for each CWE.

\paragraph{\textbf{Phase 3: Generating Chain-of-Thought Reasoning}}

In Phase 3, we build upon the outputs generated in Phases 1 and 2 to create additional content that contributes to the construction of the final prompt. Specifically, we take the LLM-generated instructions from Phase 2 and combine them with the manually curated pairwise code samples from Phase 1, which include both the vulnerable and fixed synthetic examples. These combined inputs are then used to prompt the LLM once more, guiding it to analyze the provided examples in depth. 

The primary goal is to produce a detailed chain-of-thought explanation for each pair, focusing on two key aspects: first, how the vulnerable function from Phase 1 is deemed unsafe based on the instructions generated in Phase 2, and second, how the fixed function from Phase 1 resolves the identified issues and is considered safe according to the same instructions. Note the inputs and outputs specific to this phase:

\begin{tcolorbox}[title=Phase 3: Input(s) and Output(s), coltitle=white, fonttitle=\bfseries]
\textbf{Input}: Phase 2 result (LLM-generated instructions); Phase 1 result (pairwise examples: vulnerable and fixed); Prompt

\textbf{Output}: Chain-of-thought explanations for the pairwise examples
\end{tcolorbox}

An example of the prompt used in this phase:
\begin{tcolorbox}[title=Phase 3 Prompt,  colframe=blue!50!black,colback=white, colbacktitle=blue!50!black, coltext=black!80!blue, fonttitle=\bfseries] 
You are a security expert in vulnerability detection. You specialize in identifying \{CWE-ID\}. You are given a vulnerable function and a fixed function, as well as a set of instructions that can be used to identify \{CWE-ID\}.

\newblock{Vulnerable: \{vulnerable function from Phase 1\} }

\newblock{Fixed: \{fixed function function from Phase 1\}}

Given the above functions and the 
following instructions:  \{ instructions from Phase 2 \}

1) Illustrate how the vulnerable function is vulnerable based on the instructions. 

2) Illustrate how the fixed function is non-vulnerable based on the instructions.
\end{tcolorbox}

The final output of this phase will be a detailed chain-of-thought explanation for the synthetic pair, which will serve as a crucial component in constructing the final prompt.

\paragraph{\textbf{Phase 4: Final Contrastive Chain-of-Thought Template Generation}}
\newblock{
\begin{tcolorbox}[title=Phase 4: Input(s) and Output(s), coltitle=white, fonttitle=\bfseries]
\textbf{Input}: Phase 2 result (LLM-generated instructions); Phase 1 result (pairwise examples); Phase 3 result (chain-of-thought explanations for pairwise examples); final instruction to detect real-world sample

\textbf{Output}: Final contrastive chain-of-thought template
\end{tcolorbox}
}

In the final phase, we consolidate the results by manually synthesizing a contrastive chain-of-thought template that highlights the differences between the vulnerable and fixed versions of the code. This template combines the generated instructions, the chain-of-thought for each code pair, and the reasoning process used by the LLM from each phase, restructured in a Q-and-A format. Finally, we append the instructions to identify the real-world vulnerability example.

By the end, we obtain a generalized template for a particular CWE, developed through the iteration through Phases 1-4. This process is repeated for each CWE and for each setting within the NL+CoT category. It is important to note that, while the process for NL-only settings is not detailed here, it is straightforward: the NL-only category template-generation process excludes Phases 1 and 3, as it involves generating instructions without requiring explanations of the synthetic examples. The final result for an NL+CoT template, after completing Phases 1-4, will resemble something like the template shown in Figure \ref{fig:fig3} (general template). The appendices include fully populated templates for various settings. 
\subsection{Juliet C/C++ Synthetic Examples}
\label{sec:juliet_samples}

Two C/C++ synthetic samples from the Juliet Dataset are given below. 

\begin{lstlisting}[basicstyle=\footnotesize\ttfamily, breaklines=true, breakatwhitespace=true, columns=fullflexible, keepspaces=true, frame=single, language=C, xleftmargin=0.05\textwidth, xrightmargin=0.05\textwidth, caption={Positive Synthetic Sample (Vulnerable)}]
void CWE78_OS_Command_Injection__
wchar_t_file_w32_execv_53_bad()
{
   wchar_t * data;
   wchar_t dataBuffer[100] = COMMAND_ARG2;
   data = dataBuffer;
   {
       /* Read input from a file */
       size_t dataLen = wcslen(data);
       FILE * pFile;
       /* if there is room in data, attempt to read the input from a file */
       if (100-dataLen > 1){
           pFile = fopen(FILENAME, "r");
           if (pFile != NULL)
           {
               /* POTENTIAL FLAW: Read data from a file */
               if (fgetws(data+dataLen, (int)(100-dataLen), pFile) == NULL)
               {
                   printLine("fgetws() failed");
                   /* Restore NUL terminator if fgetws fails */
                   data[dataLen] = L'\0';
               }
               fclose(pFile);
           }
       }
   }
   CWE78_OS_Command_Injection
   __wchar_t_file_w32_execv
   _53b_badSink(data);
}

\end{lstlisting}

\begin{lstlisting}[basicstyle=\footnotesize\ttfamily, breaklines=true, breakatwhitespace=true, columns=fullflexible, keepspaces=true, frame=single, language=C, xleftmargin=0.05\textwidth, xrightmargin=0.05\textwidth, caption={Negative Synthetic Sample (Non-vulnerable)}]
static void goodG2B()
{
   wchar_t * data;
   wchar_t dataBuffer[100] = COMMAND_ARG2;
   data = dataBuffer;
   /* FIX: Append a fixed string to data (not user / external input) */
   wcscat(data, L"*.*");
   CWE78_OS_Command_Injection__
    wchar_t_file_w32_execv_
    53b_goodG2BSink(data);
}
\end{lstlisting}
\subsection{Metrics}\label{sec:join}

\newcommand{\cmark}{\ding{51}}
\newcommand{\xmark}{\ding{55}}

\definecolor{pairgray}{RGB}{217,217,217}
\definecolor{pairyellow}{RGB}{255,229,153}

\definecolor{lightgray}{gray}{0.9}

\begin{table}[h]
\centering
\renewcommand{\arraystretch}{1.2}
\setlength{\tabcolsep}{6pt}
\begin{tabular}{c l l c c}
\hline
\textbf{Pair} & \textbf{Ground Truth Label} & \textbf{Prediction} &
\textbf{Row Correct?} & \makecell{\textbf{Pairwise}\\\textbf{Accurate?}} \\
\hline
\multirow{2}{*}{Pair 1} &
Vulnerable     & Vulnerable     & \cmark &
\multirow{2}{*}{\xmark} \\
\cline{2-4}
& Non-vulnerable & Vulnerable & \xmark \\
\hline
\multirow{2}{*}{Pair 2} &
Vulnerable     & Non-vulnerable & \xmark &
\multirow{2}{*}{\xmark} \\
\cline{2-4}
& Non-vulnerable & Non-vulnerable & \cmark \\
\hline
\multirow{2}{*}{Pair 3} &
Vulnerable     & Vulnerable     & \cmark &
\multirow{2}{*}{\cmark} \\
\cline{2-4}
& Non-vulnerable & Non-vulnerable & \cmark \\
\hline
\end{tabular}
\caption{Evaluation of Pairwise Accuracy. Each pair contains a \textit{vulnerable} and a \textit{non-vulnerable} sample with corresponding predictions. A row is marked correct if the prediction matches the ground truth. A pair is considered \textbf{pairwise accurate} only if \textbf{both} the vulnerable and non-vulnerable samples are predicted correctly.}
\label{tab:pairwise-acc}
\end{table}

To evaluate our study, we assess the performance of our proposed methodology using established metrics commonly applied to vulnerability detection:
Accuracy and F1 scores are standard metrics.
\[Acc=\frac{tp+tn}{tp+tn+fp+fn}\quad \scriptstyle\]
\[F1=\frac{2tp}{2tp+fp+fn}\quad  \scriptstyle\]

However, Ding et al. ~\cite{ding2024vulnerability} point out a significant limitation of these metrics: in real-world applications, the majority of code is non-vulnerable, leading to a potential bias where models achieve high accuracy simply by predicting code as \textit{non-vulnerable}. On the other hand, the F1 score, which is a harmonic mean between both precision and recall, is regarded as more suitable for imbalanced datasets. Yet it also fails to reflect the asymmetry in its penalty. Thus, they use  an alternate metric \textit{pairwise accuracy} as the most appropriate metric.

\[Pairwise Accuracy (pAcc)=\frac{Correct Pairs}{Number of Pairs}\quad  \scriptstyle\]
Pairwise accuracy is a measure of the model's ability to correctly predict the ground-truth labels for both elements of a vulnerable and fixed pair / total \# of pairs. Table~\ref{tab:pairwise-acc} illustrates pairwise accuracy using example pairs, each comprising a vulnerable and a non-vulnerable instance.
Thus, while we report and demonstrate improvement across all metrics, we particularly emphasize the success of our prompting strategies in enhancing pairwise accuracy. This focus highlights our approach's ability to better address the balance between detecting vulnerabilities accurately (and thus preventing attacks) and minimizing false alarms, enhancing our methodologies' applicability to the real world.

\section{Experimental Setup}

\label{sec:experiments}

For our main experiments, we use GPT-4 to generate our vulnerability instructions and reasoning. We evaluate the prompts and report results on the following:

\subsection{Models}

We evaluate our methodologies using state‐of‐the‐art language models. Our primary models are \textbf{OpenAI's o1 reasoning model} \textit{(o1-2024-12-17)}, \textbf{Claude} (\textit{3.5-sonnet-v2}, temperature = 0.5), and \textbf{DeepSeek-R1} \texttt{(671B parameter model}, temperature = 0.6). Additionally, we include results from models in competing work and our manual analysis: \textbf{GPT-4o} \textit{(gpt-4o-2024-05-13)}, \textbf{GPT-4} \textit{(gpt-4-turbo-2024-04-09)}, and \textbf{GPT-3.5} \textit{(gpt-3.5-turbo-0125)} \footnote{These models were evaluated on a representative subset of the full SVEN dataset, as well as on the full C/C++ dataset of CVEFixes.}. For these, we set the temperature to 0.7 and the maximum token count to 4096. All models use \texttt{top\_p = 1} and the Chat Completions API \footnote{\url{https://platform.openai.com/docs/guides/completions}}, with all other settings at their defaults. 

\subsection{Static Analyzers}
We used Amazon CodeGuru Security~\cite{codeguru}, Semgrep~\cite{semgrep}, CodeQL~\cite{codeql}, and SonarQube~\cite{sonarqube} with their default settings, all of which support static analysis on partial code. For each tool, we applied all available queries capable of operating on partial code written in either C/C++ or Python, depending on the language relevant to each CWE. In cases where no suitable queries for partial code were available for a given CWE or language, the tool was excluded from that specific evaluation. Each of the static analysis tools returns findings with information such as location, severity, and explanation of the vulnerability if detected in the code. For evaluation purposes, we consider a sample to be a true positive (TP) as long as the tool correctly reports the vulnerable location and maps it to the appropriate CWE.

\subsection{Datasets}
Previous work has shown that high-quality and reliable datasets for vulnerability detection are notably sparse and limited. Specifically, datasets like CVEFixes exhibit issues of data duplication and label inaccuracy. Nevertheless, they are frequently used in benchmark evaluations by the related work. Consequently, our study also incorporates CVEFixes to facilitate comparative analysis of our prompting techniques. We direct our efforts on demonstrating performance on novel \textbf{high-quality, real-world} datasets, specifically SVEN ~\cite{he2023large}.
\begin{table}[h]
\centering
\caption{Dataset Split}
\scriptsize
\begin{tabular}{l|l|l|l|l}
\toprule
\textbf{Dataset} & \textbf{Total} & \textbf{Size} & \textbf{\begin{tabular}[c]{@{}l@{}}Vulnerable/\\ Non-vulnerable\end{tabular}} & \textbf{CWE} \\
\midrule
\multirow{4}{*}{SVEN}      & \multirow{4}{*}{520}  & 190  & 95/95 & CWE-78       \\ \cmidrule{3-5}
                           &                       & 76  & 38/38 & CWE-190      \\ \cmidrule{3-5}
                           &                       & 114  & 57/57 & CWE-416      \\ \cmidrule{3-5}
                           &                       & 140  & 70/70 & CWE-476      \\ \midrule
\multirow{4}{*}{CVEFixes}  & \multirow{4}{*}{1784} & 500 & 152/348 & CWE-78       \\ \cmidrule{3-5}
                           &                       & 500 & 263/237 & CWE-190      \\ \cmidrule{3-5}
                           &                       & 354 & 152/202 & CWE-416      \\ \cmidrule{3-5}
                           &                       & 430 & 187/243 & CWE-476      \\ 
\bottomrule
\end{tabular}
\label{tab:ds_split}
\end{table}

\begin{itemize}[leftmargin=*]
\setlength{\itemsep}{0pt}
\setlength{\parskip}{0pt}

    \item \textbf{SVEN} \textit{(real-world)}: a manually-labeled, balanced dataset known to have 94\% label accuracy. This dataset originally comprises of 803 vulnerable/non-vulnerable pairs (1.6k samples total). We filter for the 4 CWE's, extracting 520 samples total (the full dataset for the 4 CWEs).
    \item \textbf{CVEFixes} \textit{(real-world)}: Bhandari et al. ~\cite{bhandari2021cvefixes} retrieved samples from open-source repositories using an automated collection tool. The dataset contains 39.8k samples of both the vulnerable and non-vulnerable class. We filter for our 4 CWE's of interest. We experiment on 1,784 samples total. See Table \ref{tab:ds_split}.
    \item  \textbf{Juliet C/C++} \textit{(synthetic)}: which contains test cases that cover a wide range of CWEs. For this dataset, we extract a vulnerable and fixed pair for each of the 4 CWEs of interest. See \ref{sec:juliet_samples} for samples.
\end{itemize}

\subsection{Baselines}
\textbf{Prompting Baseline:} We test using a basic template to identify if a SVEN sample is vulnerable, excluding synthetic data and NL augmentations.  The vanilla prompt performance reflects the minimum expected accuracy for the task and provides a clear reference point to evaluate the improvements achieved by our prompting strategies. 

\textbf{Static Analyzer Baseline:} We run all the SVEN samples through a static analyzer that flags vulnerable lines and reports associated CWEs/violations for each sample.

\definecolor{posgreen}{RGB}{209,239,205}
\newcommand{\pos}[1]{\cellcolor{posgreen}#1}
\newcommand{\good}[1]{\cellcolor{posgreen}#1} 

\begin{table*}[t]
\centering
\caption{\textbf{Top-performing settings.} Summary of the top-performing settings and models based on their effectiveness in pairwise accuracies for each CWE within the SVEN dataset. Results are averaged over 3 trials. The delta (\(\Delta\)) values highlight improvements relative to: the baseline prompt (basic prompt B) and the static analyzer. \colorbox{posgreen}{\textbf{Green shading}} indicates improvements (i.e., positive gains in pairwise-accuracy/accuracy/F1, or reductions in FNR/FPR).}

\label{tab:top-set}
\small
\setlength{\tabcolsep}{2pt}
\renewcommand{\arraystretch}{1.05}

\rowcolors{2}{gray!12}{white}
\begin{tabular}{l|l|l|c|c|c|c|c|c|c|c|c|c|c|c}
\toprule
\makecell{\textbf{CWE}} & \makecell{\textbf{Setting}} & \makecell{\textbf{Model}} &
\makecell{\textbf{Pairwise}\\\textbf{Acc}} &
\makecell{\textbf{\(\Delta\)}\\\textbf{Baseline}\\\textbf{Prompt}\\\textbf{Pairwise}\\\textbf{Acc}} &
\makecell{\textbf{\(\Delta\)}\\\textbf{Static}\\\textbf{Analyzer}\\\textbf{Pairwise}\\\textbf{Acc}} &
\textbf{Acc} &
\makecell{\textbf{\(\Delta\)}\\\textbf{Baseline}\\\textbf{Prompt}\\\textbf{Acc}} &
\makecell{\textbf{\(\Delta\)}\\\textbf{Static}\\\textbf{Analyzer}\\\textbf{Acc}} &
\textbf{F1} &
\makecell{\textbf{\(\Delta\)}\\\textbf{Baseline}\\\textbf{Prompt}\\\textbf{F1}} &
\makecell{\textbf{\(\Delta\)}\\\textbf{Static}\\\textbf{Analyzer}\\\textbf{F1}} &
\makecell{\textbf{FNR}\\\textbf{/ FPR}} &
\makecell{\textbf{\(\Delta\)}\\\textbf{Baseline}\\\textbf{Prompt}\\\textbf{FNR, FPR}} &
\makecell{\textbf{\(\Delta\)}\\\textbf{Static}\\\textbf{Analyzer}\\\textbf{FNR}} \\
\midrule
\makecell{CWE\\078} & \makecell{NL\\(S1)}       & OpenAI-o1   & 60.4 & \pos{+11.3} & \pos{+60.4} & 80.0 & \pos{+5.6} & \pos{+31.6} & 21.3 & -7.7  & \pos{+10.4} & 39.6 / 39.6 & \good{-11.3} & \good{-31.3} \\
\makecell{CWE\\190} & \makecell{NL+CoT\\(S0)}   & DeepSeek-R1 & 37.7 & \pos{+8.8}  & \pos{+37.7} & 65.8 & \pos{+8.8} & \pos{+15.8} & 71.7 & \pos{+26.7} & \pos{+71.7} & 62.4 / 62.4 & \good{-8.8}  & \good{-37.6} \\
\makecell{CWE\\476} & \makecell{NL\\(S0)}       & DeepSeek-R1 & 32.4 & \pos{+11.6} & \pos{+31.0} & 62.8 & \pos{+8.5} & \pos{+12.1} & 49.3 & \pos{+4.6}  & \pos{+46.5} & 67.6 / 67.6 & \good{-11.6} & \good{-31.0} \\
\makecell{CWE\\416} & \makecell{NL\\\hphantom{---}(S2)\hphantom{---}} & DeepSeek-R1 & 26.2 & \pos{+11.9} & \pos{+26.2} & 59.8 & \pos{+8.6} & \pos{+9.8}  & 38.0 & -1.7  & \pos{+38.0} & 73.9 / 73.9 & \good{-11.9} & \good{-24.7} \\
\bottomrule
\end{tabular}

\end{table*}

\section{Evaluation}
\label{sec:results}
 \noindent{\bf RQ1:} \textbf{Which prompting strategy performs best?} We investigate how different settings and models perform on our metrics.\\
  \noindent{\bf RQ2:} \textbf{How do static analyzers compare to LLMs?} \\
 \noindent{\bf RQ3:} \textbf{How do our methods compare to other relevant work?} \\
 \noindent{\bf RQ4:} \textbf{What are the strengths and limitations of LLMs in detecting software vulnerabilities using our prompting techniques?}  We analyze manual samples to uncover why different prompting strategies and models vary in effectiveness across CWEs.

\subsection{RQ1: Prompting Strategies}

\newcommand{\shiftincludegraphics}[3][-]{%
  \begin{adjustwidth}{#2}{-#2}\includegraphics[width=#1]{#3}\end{adjustwidth}%
}

\begin{figure}[b!]
\centering
\shiftincludegraphics[0.61\textwidth]{-17mm}{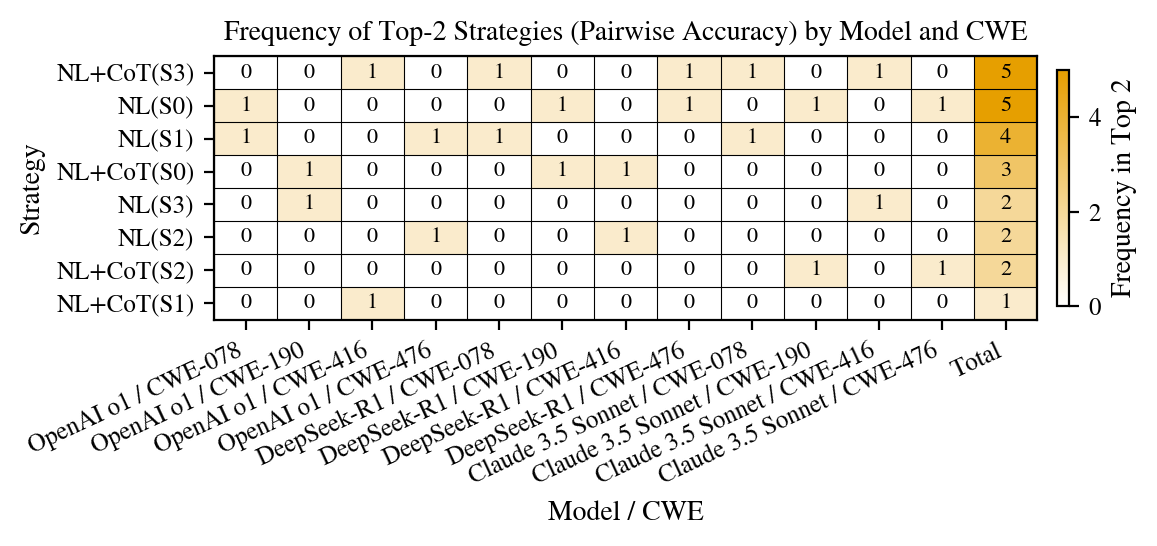}
\caption{Heatmap illustrating how often each prompting strategy appears in the top two performers across LLM–CWE combinations, based on pairwise accuracy.}
\label{fig:top-2-freq}
\end{figure}

\begin{figure}[htb]
\centering
\shiftincludegraphics[0.58\textwidth]{0mm}{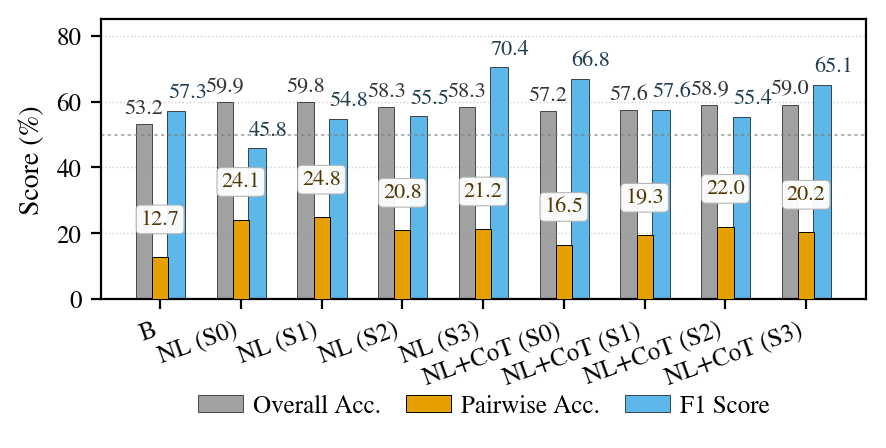}
\caption{Overall performance of our techniques across all models and all CWEs. Accuracy, F1, and pairwise accuracy reported}
\label{fig:sven_GPT_all}
\end{figure}

\noindent
 Our findings reveal that for each CWE there exists an LLM prompting strategy that significantly improves vulnerability detection compared to both the static analyzer and the prompting baseline; \textbf{thus, prompts are effective alternatives to laborious manually-crafted rule-based static analysis}. For instance, as shown in Table~\ref{tab:top-set}, the o1 model with NL (S1), which leverages LLM-generated natural language instructions, is most effective for CWE-078, while the R1 model paired with generic-CoT yields the best accuracy for CWE-190. In the evaluations of CWEs 476 and 416, DeepSeek-R1 is the top-performing model; however, the optimal prompting variant differs, with NL (S2) being most effective for CWE-416 and NL (S0) for CWE-476. Notably, no single technique is universally effective: for example, contrastive CoT templates outperform other methods for CWE-078 when evaluated with Claude and DeepSeek-R1, and similarly for CWE-416 with o1, underscoring that the most effective prompting strategy depends on the specific CWE, the specific programmatic challenges it presents, how the prompt is structured, and the inherent strengths of the model used. The heatmap in Fig. \ref{fig:top-2-freq} shows which strategies consistently rank among the top-2 performers across different LLMs and CWE cases. The ranking is based on pairwise accuracy scores. Different strategies are effective for different LLM/CWE combinations, highlighting that no one strategy universally dominates, though some (e.g. NL + CoT (S3)) tend to perform more reliably across LLM/CWE combinations.

After averaging performance across all models, CWEs, and trials, our results indicate that the NL (S1) strategy achieves the best overall pairwise accuracy, with all NL and NL+CoT approaches outperforming both the basic prompt and the generic-CoT prompt (see Figure~\ref{fig:sven_GPT_all}).

\subsection{RQ2: Comparison with Static Analyzer} Our results on static analysis tools reveal a critical trade-off in SA tools\footnote{CodeGuru, Semgrep, SonarQube, and CodeQL are designed to handle partial code. We utilize them with their default settings, wherein built-in CWE-specific rules are invoked by default.}: even when these tools strive for high precision, ensuring that any flagged vulnerability is indeed a true vulnerability, they suffer from extremely low recall in the partial code setting, thereby missing a significant number of actual vulnerabilities.
CWE-078 exhibits a precision of 40\% but a recall of only 6.3\%, leading to an F1 score of approximately 11\% and a pairwise accuracy of 0. Similarly, for CWE-476, although the tool achieves 100\% precision when it does flag a vulnerability, it only detects about 1.4\% of actual cases with a pairwise accuracy of 1.4\%. In contrast, for CWE-190 and CWE-416, the overall accuracy of 50\% essentially reflects the fact that the tool often defaults to a negative prediction, failing to identify any true vulnerabilities. Moreover, pairwise accuracies remain close to zero for all CWEs, with the highest observed value only around 3\%. \textbf{Our CWE-specific prompts are much more effective.}

\begin{table*}[h]
\centering
\caption{\textbf{Static Analysis Results} after running several SA tools on our 4 CWEs.}
\label{tab:satools}
\resizebox{\textwidth}{!}{%
\rowcolors{2}{gray!15}{white} 
\begin{tabular}{lccccccccccccc}
\toprule
\textbf{Tool} & \textbf{CWE} & \textbf{Acc.} & \textbf{Prec.} & \textbf{Rec.} & \textbf{F1} & 
\textbf{TP} & \textbf{FP} & \textbf{TN} & \textbf{FN} & 
\textbf{FPR} & \textbf{FNR} & \textbf{Pair. Acc.} & \textbf{Total} \\
\midrule
CodeGuru   & CWE-190 & 0.50 & 0    & 0.0   & 0.00 & 0 & 0 & 38 & 38 & 0.000 & 1.000 & 0.0 & 76  \\
CodeGuru   & CWE-416 & 0.50 & 0    & 0.0   & 0.00 & 0 & 0 & 57 & 57 & 0.000 & 1.000 & 0.0 & 114 \\
CodeGuru   & CWE-078 & 0.48 & 0.4  & 0.063 & 0.11 & 6 & 9 & 86 & 89 & 0.095 & 0.937 & 0.0 & 190 \\
CodeGuru   & CWE-476 & 0.51 & 1.0  & 0.014 & 0.03 & 1 & 0 & 70 & 69 & 0.000 & 0.986 & 0.014 & 140 \\
Semgrep    & CWE-190 & 0.50 & 0    & 0.0   & 0.00 & 0 & 0 & 38 & 38 & 0.000 & 1.000 & 0.0 & 76  \\
Semgrep    & CWE-416 & 0.50 & 0    & 0.0   & 0.00 & 0 & 0 & 57 & 57 & 0.000 & 1.000 & 0.0 & 114 \\
Semgrep    & CWE-078 & 0.52 & 0.8  & 0.04  & 0.08 & 4 & 1 & 94 & 91 & 0.010 & 0.957 & 0.03 & 190 \\
Semgrep    & CWE-476 & 0.49 & 0    & 0.0   & 0.00 & 0 & 2 & 68 & 70 & 0.028 & 1.000 & 0.0 & 140 \\
SonarQube  & CWE-190 & 0.50 & 0    & 0.0   & 0.00 & 0 & 0 & 38 & 38 & 0.000 & 1.000 & 0.0 & 76  \\
SonarQube  & CWE-416 & 0.50 & 0    & 0.0   & 0.00 & 0 & 0 & 57 & 57 & 0.000 & 1.000 & 0.0 & 114 \\
SonarQube  & CWE-078 & 0.50 & 0    & 0.0   & 0.00 & 0 & 0 & 95 & 95 & 0.000 & 1.000 & 0.0 & 190 \\
SonarQube  & CWE-476 & 0.50 & 0    & 0.0   & 0.00 & 0 & 0 & 70 & 70 & 0.000 & 1.000 & 0.0 & 140 \\
CodeQL     & CWE-078 & 0.50 & 0    & 0.0   & 0.00 & 0 & 0 & 95 & 95 & 0.000 & 1.000 & 0.0 & 190 \\
\bottomrule
\end{tabular}%
}
\end{table*}

\subsection{RQ3: Comparison with SOTA Methods} The \texttt{CVEFixes} dataset spans multiple programming languages. Previous studies, namely Khare et al. \cite{khare2023understanding}, primarily focused on applying GPT-3.5 and GPT-4 to analyze the C/C++ subset of the dataset, covering our 4 CWEs of interest. To align with this study, we similarly restrict our analysis to  \texttt{CVEFixes C/C++} for our 4 CWEs of interest. We test on 1,201 samples total spanning the 4 CWEs. We replicate the studies conducted by the authors as closely as possible.\footnote{Note: deprecations on the OpenAI platform have resulted in the unavailability of certain GPT models referenced in their work. As such, we adopt the recommended replacement models ({\url{https://platform.openai.com/docs/deprecations}}) suggested by OpenAI in these cases in order to ensure continuity in our analysis.} 
We report the result from the best-performing model and prompting strategy. As shown in Table \ref{tab:model-comparison}, we show that for each CWE, there exists a strategy that achieves the best results, achieving up to a 12\% improvement in accuracy with our prompting. 

\begin{table}[H]
\centering
\caption{Comparison of method accuracies on CVEFixes C/C++.}
\scriptsize
\begin{tabular}{l|l|l|l|l}
\toprule
\textbf{Method} & \textbf{Dataset} & \textbf{CWE} & \textbf{Model} & \textbf{Acc} \\
\midrule
Khare et al.    & CVEFixes C/C++   & 78           & GPT-3.5        & 0.63         \\
Our Approach           & CVEFixes C/C++   & 78           & GPT-3.5        & \textbf{0.67}\\
Our Approach           & CVEFixes (All)   & 78           & GPT-3.5        & 0.67         \\
\midrule
Khare et al.    & CVEFixes C/C++   & 190          & GPT-4          & 0.48         \\
Our Approach           & CVEFixes C/C++   & 190          & GPT-4          & \textbf{0.60}\\
Our Approach           & CVEFixes (All)   & 190          & GPT-4          & 0.58         \\
\midrule
Khare et al.    & CVEFixes C/C++   & 416          & GPT-4          & 0.46         \\
Our Approach           & CVEFixes C/C++   & 416          & GPT-4          & \textbf{0.56}\\
Our Approach           & CVEFixes (All)   & 416          & GPT-4          & 0.59         \\
\midrule
Khare et al.    & CVEFixes C/C++   & 476          & GPT-4          & 0.58         \\
Our Approach           & CVEFixes C/C++   & 476          & GPT-4          & \textbf{0.59}\\
Our Approach           & CVEFixes (All)   & 476          & GPT-4          & 0.58         \\
\bottomrule
\end{tabular}
\label{tab:model-comparison}
\end{table}
\subsection{RQ4: Strengths and Limitations of LLMs in Identifying Vulnerabilities}

We manually reviewed 414 LLM-generated responses showing notably high or low performance, focusing on the selected instances in the table. For each instance, we analyzed 46 responses on both vulnerable and fixed versions, then classified them into four categories—\(LR, L\overline{R}, \overline{L}R, \overline{L}\overline{R}\)—based on whether the label (L) and reasoning (R) were correct or incorrect.
\begin{table}[!ht]
\centering
\caption{~\textbf{LLM Response Analysis.} We manually analyze 414 output responses by the LLM (46 for each CWE-setting-model combo).}
\resizebox{.45\textwidth}{!}{
\begin{tabular}{l|l|l|l|r|r|r}
\toprule
\textbf{CWE} & \textbf{Setting} & \textbf{Model} & \textbf{$LR$} & \textbf{$L\overline{R}$} & \textbf{$\overline{L}R$} & \textbf{$\overline{L}\overline{R}$} \\ 
\midrule
CWE-78       & NL (S1)        & OpenAI-o1 & 40 & 0  & 6  & 0  \\ 
\midrule
CWE-190       & NL+CoT (S0)        & DeepSeek-R1 & 33 & 0  & 10  & 3  \\ 
\midrule
CWE-78   & NL (S1)        & GPT-4o    & 37 & 1  & 8  & 0  \\ 
\midrule
CWE-78   & NL+CoT (S3)    & GPT-4o    & 36 & 1  & 9  & 0  \\ 
\midrule
CWE-190  & NL+CoT (S3)    & GPT-4     & 34 & 1  & 8  & 3 \\ 
\midrule
CWE-416  & NL+CoT (S1)    & GPT-4     & 25 & 0  & 0  & 21 \\ 
\midrule
CWE-190  & NL+CoT (S1)    & GPT-3.5   & 20 & 0  & 3  & 23 \\ 
\midrule
CWE-476  & NL (S0)        & DeepSeek-R1 & 20 & 9  & 2  & 15 \\ 
\midrule
CWE-476  & NL+CoT (S1)    & GPT-3.5   & 23 & 0  & 0  & 23 \\ 
\bottomrule
\end{tabular}
}
\label{tab:response-analysis}
\end{table}

\paragraph{CWE-078} As shown in Table~\ref{tab:response-analysis}, all models performed strongly in classifying CWE-78, accurately detecting unsafe command construction (e.g., user input concatenation). Notably, o1 in NL (S1) excelled at recognizing secure coding practices (e.g., argument lists, parameterization) but struggled to detect indirect user input pathways, suggesting a need for enhanced contextual understanding and data flow analysis.

\paragraph{CWE-476} While DeepSeek-R1 shows valid reasoning, its coverage of potential vulnerability sites is limited. It mainly uses a top-down analysis strategy, focusing on function inputs and early variable declarations to make broad conclusions about null pointer dereferences across the whole function, rather than examining each potential vulnerability site independently. This selective reasoning and reliance on early-stage analysis lead to suboptimal accuracy. Other LLMs like GPT-3.5 are pedantic about NULL checks on all declared variables, often failing to recognize NULL-valued pointers within data structures like structs. 

\paragraph{CWE-190} DeepSeek-R1 in NL+CoT (S0) effectively detects integer overflow vulnerabilities but sometimes overestimates risks, overlooking explicit checks already in place. It frequently flags safe memory allocation functions (e.g., kmalloc\_array, calloc, jas\_alloc2) due to insufficient size-check tracking, and misses overflows mitigated later in execution. However, its detailed explanations often pinpoint buffer limits and signed/unsigned mismatches, and other LLMs like GPT-4 also detect issues in buffer-size calculations, albeit sometimes adding unnecessary overflow checks.

\paragraph{CWE-416} Analysis of GPT-4 responses on CWE-416 samples in NL+CoT (S1) reveals limited understanding of memory management, synchronization in concurrent environments, and edge-case handling. Many missed vulnerabilities stem from missing context in the code snippets.

\section{Related Work}
 \label{sec:related}

Earlier methods for detecting vulnerabilities used convolutional neural representations ~\cite{russell2018automatedvulnerabilitydetection}, transformer models, and graph-based networks like Devign~\cite{zhou2019devign}, ReVeal~\cite{chakraborty2021deep}, IVDetect~\cite{li2021vulnerability}, and LineVD~\cite{hin2022linevd} to analyze code relationships. More recent approaches with fine-tuned transformer models, such as CodeBERT~\cite{feng2020codebert}, LineVul~\cite{fu2022linevul}, and UniXcoder~\cite{guo2022unixcoder}, have shown better performance. However, issues like data leakage, label mistakes, and duplication in older datasets~\cite{ding2024vulnerability} have led to efforts to create cleaner and more reliable datasets~\cite{ding2024vulnerability, he2023large}. Even with these improvements, LLMs still struggle to detect vulnerabilities in zero-shot settings~\cite{steenhoek2024comprehensive, ding2024vulnerability}.

Other works leverage LLMs for vulnerability detection or utilize CWE-specific designs, many employing LLMs for generating useful representations, fine-tuning per CWE type, using older transformer models as base models, or continuing to evaluate on problematic datasets.

Du et al. ~\cite{du2024generalizationenhancedcodevulnerabilitydetection} leverage GPT-4 to generate interpretations of vulnerabilities and show it is effective in instruction-tuning vulnerability detection models; however, they still target problematic datasets.  Lekssays et al. ~\cite{lekssays2025llmxcpgcontextawarevulnerabilitydetection} leverage LLMs to generate queries for traversing code-property graphs. Gao et al. ~\cite{gao2023fargonevulnerabilitydetection} compare LLMs against graph-based approaches, and show that several LLMs outperform traditional CNN/Graph-based models. Cao et al. ~\cite{cao-etal-2024-realvul} focus specifically on PHP vulnerabilities and have to fine-tune their model for each type of CWE, incurring substantial overhead. Atiiq et al. ~\cite{atiiq2024generalistspecialistexploringcwespecific} and Yang et al. ~\cite{oneforall} advocate against “one-for-all” designs, such as those that train a single model to handle multiple vulnerabilities or consider multiple CWE types at once. Rather than handling multiple vulnerability types in one model, ~\cite{atiiq2024generalistspecialistexploringcwespecific} train CWE-specific classifiers, while ~\cite{oneforall} use a mixture-of-experts (MoE) approach, clustering CWEs that share similar vulnerability patterns and training “experts”, which are designed to handle a particular CWE (or cominbation of CWEs), over graph and transformer architectures as base models. 
Ji et al. ~\cite{ji-etal-2024-applying} also group CWEs that share similar patterns by taking into account vector representations of the vulnerabilities and applying contrastive learning to bring related CWEs closer together.  Another work by Fu et al. ~\cite{fu2023chatgptvulnerabilitydetectionclassification} found that domain-specific prompting is still necessary to make LLMs effective in vulnerability detection. 
The direction of these works highlight the gap and necessity in exploring effective CWE-specific approaches that integrate domain knowledge at inference time.

New prompting techniques that add domain-specific knowledge show promise on synthetic data but are still limited on real-world vulnerabilities~\cite{ullah2024llms, khare2023understanding}. Hybrid methods, like combining static analysis with LLMs~\cite{li2024llmassistedstaticanalysisdetecting}, reduce false positives but are less-applicable in the partial-code setting, requiring full project builds to operate. In modern software development settings like GitHub Copilot, where LLMs generate code incrementally (e.g., line-by-line), relying on full project builds is unrealistic. Our prompting strategies are designed to detect vulnerabilities even in partial code. This emphasizes the importance of reliable solutions for identifying vulnerabilities at every stage of development, including safeguarding LLMs from generating insecure code.

\section{Limitations}
Cost poses a significant limitation in our experiments, particularly as academic researchers aiming to utilize state-of-the-art models. We opt to use the latest SOTA open-source model at the time of writing: DeepSeek-R1, Claude 3.5 Sonnet, as well as OpenAI's o1 and GPT-based models due to their recognized quality. Throughout the course of this work, we made over 190,000 API calls to support our experiments and analyses. As a result of these constraints, we focused our investigation on 4 specific CWEs. We have provided a cost analysis; see Section \ref{sec:cost-analysis}. Consequently, we do not conduct a comprehensive cross-analysis across CWEs. As emphasized in our study, the quality of the real-world datasets at hand is critical (for this reason we targeted SVEN \cite{he2023large}). Additionally, constructing a reasoning framework using synthetic examples requires identifying CWEs that are consistently present across \textbf{both} our high-quality/real-world and synthetic datasets, which further restricts our experimental scope.

There may be variations of prompt design where formatting has a greater impact on prompt strategy performance. One such approach is tag-based formatting. As described in Section \ref{sec:tagbasedformatting}, we experimented with tag-based formatting and found that the relative rankings of prompt strategies remained robust despite its inclusion. However, given the large space of possible formatting styles, it remains possible that some alternative formatting styles could prove optimal. A comprehensive exploration of formatting styles, however, is beyond the scope of this work.
\subsection{Tag-Based Formatting} 
\label{sec:tagbasedformatting}
The verbosity instruction was added only to the basic prompt; this was implemented to avoid inconclusive responses, which would've further degraded baseline performance. 
We repeat the basic prompt experiments– this time without the verbosity instruction and with incorporating the $<answer>...</answer>$ tags. 
Overall, we find: in most cases, adding the tags reduced pairwise accuracy (pAcc) compared to our version of the basic prompt with the verbosity instruction; in the few cases where the answer tag improved pAcc, the difference was insignificant ($<$1.5 $\%$)
The performance of our Basic Prompt (Setting 1) vs Basic Prompt + without verbosity instruction + with $<answer>$ tags (Setting 2) is not significantly different enough to affect the rankings of the strategies. See Table \ref{tab:tagexp}. The ranking of our strategies remains robust; the tags do not affect the outcome of the best performing strategy shown in Table \ref{tab:top-set}.
\begin{table}[t]
\centering
\caption{\textbf{Tag-based formatting experiment} after re-running the basic prompt without the verbosity instruction and with tag-based formatting. S1 refers to Setting 1; S2 refers to Setting 2}
\label{tab:tagexp}
{\Large
\resizebox{\linewidth}{!}{%
\begin{tabular}{l|l|c|c|c|c|c|c}
\toprule
\textbf{CWE} & \textbf{Model} & \textbf{S1 F1} & \textbf{S1 Acc} & \textbf{S1 pAcc} & \textbf{S2 F1} & \textbf{S2 Acc} & \textbf{S2 pAcc} \\
\midrule
\rowcolor{gray!10}
78  & R1  & 53.0  & 67.2  & 41.1  & 63.3  & 65.1  & 34.3  \\
78  & o1  & 29.0  & 74.4  & 49.1  & 28.0  & 74.7  & 50.5  \\
\rowcolor{gray!10}
190 & R1  & 45.0  & 57.0  & 28.9  & 53.3  & 59.6  & 28.9  \\
190 & o1  & 26.3  & 56.9  & 26.3  & 16.3  & 59.6  & 23.6  \\
\rowcolor{gray!10}
416 & R1  & 39.7  & 51.2  & 14.3  & 44.0  & 53.5  & 15.8  \\
416 & o1  & 24.0  & 49.7  & 4.2   & 10.6  & 40.7  & 4.0   \\
\rowcolor{gray!10}
476 & R1  & 44.7  & 54.3  & 19.8  & 62.0  & 50.4  & 18.2  \\
476 & o1  & 24.3  & 51.9  & 7.2   & 24.0  & 52.0  & 8.0   \\
\bottomrule
\end{tabular}%
}
}
\end{table}

\subsection{Cost Analysis}
\label{sec:cost-analysis}

\newcommand{\twocell}[2]{%
  \begingroup\setlength{\fboxsep}{1pt}%
  \makecell{%
    \colorbox{gray!12}{\strut #1}\\[-0.2ex]%
    \colorbox{white}{\strut #2}%
  }%
  \endgroup
}
\newcommand{\twocellcwe}[1]{%
  \begingroup\setlength{\fboxsep}{1pt}%
  \makecell{%
    \colorbox{gray!12}{\strut #1}\\[-0.2ex]%
    \colorbox{white}{\strut\phantom{#1}}%
  }%
  \endgroup
}

\begin{table}[t]
\caption{Cost analysis.}
\centering
\scriptsize
\setlength{\tabcolsep}{3pt}
\renewcommand{\arraystretch}{0.95}
\begin{tabular}{l|c|r|r|r|r|r|r|r|r} 
\toprule

\textbf{CWE} & \makecell{\textbf{S}} & \makecell{\textbf{B}} & \makecell{\textbf{NL}\\\textbf{S0}}&
\makecell{\textbf{NL}\\\textbf{S1}} & \makecell{\textbf{NL}\\\textbf{S2}} & \makecell{\textbf{NL}\\\textbf{S3}} &
\makecell{\textbf{NL+CoT}\\\textbf{S1}} & \makecell{\textbf{NL+CoT}\\\textbf{S2}} & \makecell{\textbf{NL+CoT}\\\textbf{S3}} \\
\midrule
\twocellcwe{CWE-078} & \twocell{1}{2} & \twocell{317}{312} & \twocell{325}{320}
  & \twocell{716}{711} & \twocell{489}{484} & \twocell{831}{826}
  & \twocell{1319}{1314} & \twocell{1190}{1185} & \twocell{1626}{1621} \\
\twocellcwe{CWE-190} & \twocell{1}{2} & \twocell{1439}{1185} & \twocell{1449}{1195}
  & \twocell{1886}{1632} & \twocell{1979}{1725} & \twocell{1636}{1382}
  & \twocell{2413}{2159} & \twocell{2717}{2463} & \twocell{2592}{2338} \\
\twocellcwe{CWE-416} & \twocell{1}{2} & \twocell{577}{996} & \twocell{585}{1004}
  & \twocell{870}{1289} & \twocell{829}{1248} & \twocell{833}{1252}
  & \twocell{1812}{2231} & \twocell{1860}{2279} & \twocell{1663}{2082} \\
\twocellcwe{CWE-476} & \twocell{1}{2} & \twocell{1363}{1565} & \twocell{1372}{1574}
  & \twocell{1806}{2008} & \twocell{1611}{1813} & \twocell{1440}{1642}
  & \twocell{2598}{2800} & \twocell{2601}{2803} & \twocell{2685}{2887} \\
\bottomrule
\end{tabular}

\end{table}
\label{tab:cost}

Table~\ref{tab:cost} provides the average token count for each experiment using the gpt-3.5-turbo tokenizer. An experiment in setting 1 means that it is carried out using 23 samples; an experiment in setting 2 means that it is carried out using all available samples.

On average, the length of the prompts stay within the context window of language models. Most of the tokens come from the original code excerpt, which can vary in length. The additional cost of our prompting strategy does not turn out to be a major cost overhead.
\section{Conclusions}
\vspace{-1mm}
We demonstrate the effectiveness of CWE-specific LLM prompts over static analysis rules in the partial code setting. These prompting strategies not only boost traditional metrics like accuracy, F1, FPR, and FNR, but also significantly improve \emph{pairwise accuracy}, a critical indicator of a system's ability to identify true vulnerabilities while mitigating false alarms. 
In modern workflows where developers often (1) reuse existing implementations (2) write and review partial code, and (3) interact with LLM-systems in snippet-fashion, vulnerability detection in the partial code setting is critical. Furthermore and just as importantly, attackers manipulate partial code for malicious purposes, using it to evade detection and stress-test LLM defenses, which can induce insecure outputs in the LLMs. This highlights the relevance and challenge of detecting vulnerabilities in partial-code generation. Our method requires \emph{no model training or project build}, works on snippets, and complements static analysis by cutting false negatives while catching true positives at review time. To facilitate reproducibility, we provide a replication package containing source code and datasets.\footnote{\url{https://github.com/SoftwEngLab/partial-code-analysis-llm-cwe}}

\section{Acknowledgements}

This work was supported in part by the National Science Foundation Graduate Research Fellowship Program (NSF GRFP), NSF CNS-2247370, NSF CCF-2313055, 
DARPA/NIWC Pacific N66001-21-C-4018. Any opinions, findings,
conclusions or recommendations expressed herein are those of the authors and do not necessarily reflect those of the US Government, NSF, or DARPA.

\bibliography{bare_jrnl}
\bibliographystyle{IEEEtran}

\lstset{
  basicstyle=\footnotesize\ttfamily,
  breaklines=true,
  breakatwhitespace=true,
  columns=fullflexible,
  keepspaces=true,
  frame=single,
  language=C,
  xleftmargin=0.05\textwidth,
  xrightmargin=0.05\textwidth
}

\tcbset{
  promptstyle/.style={
    breakable,         
    enhanced,           
    colback=gray!10,
    colframe=white,
    boxrule=0pt,
    arc=0pt,
    left=6pt,
    right=6pt,
    top=4pt,
    bottom=4pt
  }
}

\section{Appendix}
\subsection{NL+CoT (S0) Sample Prompt for (CWE-190)}
\label{cot-s0-example}

\begin{tcolorbox}[promptstyle]
You are a security expert. You will analyze new examples for vulnerabilities.

Use the following instructions:

The product performs a calculation that can produce an integer overflow or wraparound when the logic assumes that the resulting value will always be larger than the original value. This occurs when an integer value is incremented to a value that is too large to store in the associated representation. When this occurs, the value may become a very small or negative number.

Q: Check for any vulnerable sections of code in the new example, then determine whether the new example is vulnerable or non-vulnerable based on the instructions.

\texttt{<EXAMPLE START>}
\begin{lstlisting}
nresp = packet_get_int();
if (nresp > 0) {
  response = xmalloc(nresp*sizeof(char*));
  for (i = 0; i < nresp; i++) response[i] = packet_get_string(NULL);
}
\end{lstlisting}
\texttt{<EXAMPLE END>}

Your answer:

The following code excerpt from OpenSSH 3.3 demonstrates a classic case of integer overflow. If nresp has the value 1073741824 and sizeof(char*) has its typical value of 4, then the result of the operation nresp*sizeof(char*) overflows, and the argument to xmalloc() will be 0. Most malloc() implementations will happily allocate a 0-byte buffer, causing the subsequent loop iterations to overflow the heap buffer response.

Q: Check for any vulnerable sections of code in the new example, then determine whether the new example is vulnerable or non-vulnerable based on the instructions.
\texttt{<NEW EXAMPLE START>}
\begin{lstlisting}
long WebPImage::getHeaderOffset(byte *data, long data_size,
                                byte *header, long header_size) {
  long pos = -1;
  for (long i=0; i < data_size - header_size; i++) {
    if (memcmp(header, &data[i], header_size) == 0) {
      pos = i;
      break;
    }
  }
  return pos;
}
\end{lstlisting}
\texttt{<NEW EXAMPLE END>}

Your answer:
\end{tcolorbox}

\subsection{NL+CoT (S1) Sample Prompt (CWE-190)}

\begin{tcolorbox}[promptstyle]

\label{cot-s1-example}
You are a security expert. You will analyze new examples for vulnerabilities.

Use the following tests:
\begin{itemize}
\setlength{\itemsep}{0pt}
\setlength{\parskip}{0pt}
    \item \textbf{Test 1 -} Ensures that arithmetic operations involving integers adhere to the limits of the data type, specifically testing operations at, just below, and just above the maximum and minimum values. This is essential to prevent overflow or underflow, which could lead to unexpected behavior or vulnerabilities.
    \item \textbf{Test 2 -} Ensures that the selection of data types for integers is appropriate for their intended use, with a particular focus on choosing wider types (e.g., int64\_t over int32\_t) when the risk of overflow is significant, and the correct use of signed versus unsigned integers based on the context.
    \item \textbf{Test 3 -} Verifies that the example utilizes libraries designed for safe arithmetic operations, such as SafeInt in C++, which can prevent overflows and underflows by throwing exceptions or signaling errors. This test checks for both the presence of these libraries in the codebase and their correct application in arithmetic operations.
    \item \textbf{Test 4 -} Confirms that, in the absence of safe arithmetic libraries, the code manually checks for potential overflow and underflow conditions before performing arithmetic operations. This involves validating that arithmetic results remain within the bounds of the data types used.
    \item \textbf{Test 5 -} Ensures all inputs, whether user-supplied or externally generated, that are used in arithmetic operations undergo stringent validation checks. This includes confirming that inputs are within expected ranges and are of the correct format and type to prevent overflows or unexpected behavior.
    \item \textbf{Test 6 -} Focuses on verifying that explicit range checks are implemented for all inputs prior to their use in arithmetic operations. This step is crucial for ensuring that inputs do not lead to integer overflow or wraparound by exceeding the operational bounds of the data types involved.
\end{itemize}

Q: Check for any vulnerable sections of code in the new example, then determine whether the new example is vulnerable or non-vulnerable based on the tests.

\texttt{<EXAMPLE START>}
\begin{lstlisting}[basicstyle=\footnotesize\ttfamily, breaklines=true, breakatwhitespace=true, columns=fullflexible, keepspaces=true, frame=single, language=C, xleftmargin=0.05\textwidth, xrightmargin=0.05\textwidth]
void CWE190_Integer_Overflow__int64_t
_fscanf_multiply_81_bad
::action(int64_t data) const
{
   if(data > 0) /* ensure we won't have an underflow */
   {
       /* POTENTIAL FLAW: if (data*2) > LLONG_MAX, this will overflow */
       int64_t result = data * 2;
       printLongLongLine(result);
   }
}
\end{lstlisting}
\texttt{<EXAMPLE END>}

Your answer:
\begin{itemize}
    \item \textbf{Test 1} - Fail. The vulnerable function fails this test because it directly performs the arithmetic operation (data * 2) without first checking if the result would exceed LLONG\_MAX, the maximum value for an int64\_t. This can lead to an integer overflow if data is greater than half of LLONG\_MAX.
    \item \textbf{Test 2} - Pass. This function correctly uses an int64\_t data type for data, which is suitable for handling large integers. However, the choice of data type alone does not prevent overflow without proper checks, as seen in the operation.
    \item \textbf{Test 3} - Fail. There is no evidence that the function uses libraries designed for safe arithmetic operations. It directly multiplies data by 2 without utilizing any mechanisms to prevent overflow, indicating a lack of safe arithmetic practices.
    \item \textbf{Test 4} - Fail. The function lacks manual checks for potential overflow conditions before performing the arithmetic operation. This is a clear failure to manually validate arithmetic results within the bounds of the int64\_t data type, leading to possible overflow.
    \item \textbf{Test 5} - Pass. While the function ensures data is greater than 0 to prevent underflow, it does not validate the upper limit, showing partial adherence to input validation. This is insufficient for preventing overflow, however.
    \item \textbf{Test 6} - Fail. Explicit range checks for data before the arithmetic operation are absent. The function only checks for underflow (data > 0) and neglects to verify that data does not cause overflow when doubled, failing this test.

\end{itemize}

Q: Check for any vulnerable sections of code in the new example, then determine whether the new example is vulnerable or non-vulnerable based on the tests.

\texttt{<EXAMPLE START>}
\begin{lstlisting}[basicstyle=\footnotesize\ttfamily, breaklines=true, breakatwhitespace=true, columns=fullflexible, keepspaces=true, frame=single, language=C, xleftmargin=0.05\textwidth, xrightmargin=0.05\textwidth]
void CWE190_Integer_Overflow__int64_t
_fscanf_multiply_81_goodB2G::
action(int64_t data) const
{
   if(data > 0) /* ensure we won't have an underflow */
   {
       /* FIX: Add a check to prevent an overflow from occurring */
       if (data < (LLONG_MAX/2))
       {
           int64_t result = data * 2;
           printLongLongLine(result);
       }
       else
       {
           printLine("data value is too large to perform arithmetic safely.");
       }
   }
}
\end{lstlisting}
\texttt{<EXAMPLE END>}

Your answer:
\begin{itemize}
    \item \textbf{Test 1} - Pass. The fixed function introduces a crucial check if (data < (LLONG\_MAX/2)) to ensure that the arithmetic operation does not cause an overflow. This adheres to the first test by confirming operations stay within the limits of the data type.
    \item \textbf{Test 2} - Pass. Similar to the vulnerable function, the fixed version uses int64\_t for data, which is appropriate for the function's requirements. The added conditional check further ensures the data type's limits are respected.
    \item \textbf{Test 3} - Pass. While the fixed function does not explicitly use a library for safe arithmetic, it manually implements a check to prevent overflow, which is the essence of what such libraries aim to achieve. Therefore, in the context of preventing integer overflow, this could be considered a conditional pass.
    \item \textbf{Test 4} - Pass. The fixed function passes this test by manually checking for potential overflow (if (data < (LLONG\_MAX/2))). This validation ensures arithmetic results remain within the bounds of the int64\_t data type.
    \item \textbf{Test 5} - Pass. The fixed function maintains the initial check for data being greater than 0 and adds an overflow protection measure. This demonstrates thorough input validation, ensuring inputs are within expected ranges before use in arithmetic operations.
    \item \textbf{Test 6} - Pass. By implementing an explicit range check (if (data < (LLONG\_MAX/2))) before the arithmetic operation, the fixed function successfully verifies that all inputs do not exceed operational bounds, effectively preventing integer overflow or wraparound.

\end{itemize}

Q: Check for any vulnerable sections of code in the new example, then determine whether the new example is vulnerable or non-vulnerable based on the tests.

\texttt{<NEW EXAMPLE START>}
\begin{lstlisting}[basicstyle=\footnotesize\ttfamily, breaklines=true, breakatwhitespace=true, columns=fullflexible, keepspaces=true, frame=single, language=C, xleftmargin=0.05\textwidth, xrightmargin=0.05\textwidth]
int perf_cpu_time
_max_percent_handler
(struct ctl_table *table, int write,
    void __user *buffer, 
    size_t *lenp, loff_t *ppos)
{
	int ret = proc_dointvec(table, write, buffer, lenp, ppos);

	if (ret || !write)
		return ret;

	if (sysctl_perf_cpu_time_max_percent == 100 ||
	    sysctl_perf_cpu_time_max_percent == 0) {
		printk(KERN_WARNING
		       "perf: Dynamic interrupt throttling disabled, can hang your system!\n");
		WRITE_ONCE(perf_sample_allowed_ns, 0);
	} else {
		update_perf_cpu_limits();
	}

	return 0;
}
\end{lstlisting}
\texttt{<NEW EXAMPLE END>}

Your answer:

\end{tcolorbox}

\subsection{NL+CoT (S3) Sample Prompt (CWE-78)}
\label{cot-s3-example}

\begin{tcolorbox}[promptstyle]
You are a security expert. You will analyze new examples for vulnerabilities.

Use the following instructions:
\begin{itemize}
\setlength{\itemsep}{0pt}
\setlength{\parskip}{0pt}
    \item \textbf{User Input Handling:} Scrutinize how user inputs are received and processed. Pay particular attention to any point where user input could be included in commands to be executed at the OS level. This includes web form inputs, URL parameters, headers, cookies, and any external source that a user can manipulate.
    \item \textbf{Command Construction:} Identify all instances where system commands are constructed and executed. Look for functions like \texttt{exec()}, \texttt{system()}, \texttt{Runtime.exec()} in Java, \texttt{ProcessBuilder} in Java, \texttt{subprocess.Popen()} in Python, or similar functions in other languages. Review how these commands are constructed, especially any concatenation operations that involve user-controlled data.
    \item \textbf{Validation and Sanitization:} Check for the presence and effectiveness of input validation and sanitization routines. Ensure that data used in command execution undergoes strict validation against a whitelist of allowed characters or patterns. Sanitization should remove or neutralize potentially malicious characters (e.g., semicolons, ampersands, and other shell metacharacters).
    \item \textbf{Use of Safe APIs:} Verify whether the code uses safer API alternatives that avoid shell interpretation, such as argument arrays in \texttt{exec()} calls instead of single string commands. These APIs are designed to reduce the risk of injection by handling each argument as separate from the command itself, thereby preventing the incorporation of shell control characters from user input.
    \item \textbf{Parameterized Commands:} For languages and frameworks that support it, check if commands are parameterized or templated in a way that separates code from data. This pattern is similar to using prepared statements in SQL and ensures that user input cannot be interpreted as part of the command.
    \item \textbf{External Libraries and Dependencies:} Review the use of external libraries or dependencies that might introduce command execution functionality. Ensure these libraries are reputable, well-maintained, and do not have known vulnerabilities that could be exploited for command injection.
\end{itemize}

Q: Check for any vulnerable sections of code in the new example, then determine whether the new example is vulnerable or non-vulnerable based on the tests.

\texttt{<EXAMPLE START>}
\begin{lstlisting}[basicstyle=\footnotesize\ttfamily, breaklines=true, breakatwhitespace=true, columns=fullflexible, keepspaces=true, frame=single, language=C, xleftmargin=0.05\textwidth, xrightmargin=0.05\textwidth]
void CWE78_OS_Command_Injection__
wchar_t_file_w32_execv_53_bad()
{
   wchar_t * data;
   wchar_t dataBuffer[100] = COMMAND_ARG2;
   data = dataBuffer;
   {
       /* Read input from a file */
       size_t dataLen = wcslen(data);
       FILE * pFile;
       /* if there is room in data, attempt to read the input from a file */
       if (100-dataLen > 1)
       {
           pFile = fopen(FILENAME, "r");
           if (pFile != NULL)
           {
               /* POTENTIAL FLAW: Read data from a file */
               if (fgetws(data+dataLen, (int)(100-dataLen), pFile) == NULL)
               {
                   printLine("fgetws() failed");
                   /* Restore NUL terminator if fgetws fails */
                   data[dataLen] = L'\0';
               }
               fclose(pFile);
           }
       }
   }
   CWE78_OS_Command_Injection
   __wchar_t_file_w32_execv
   _53b_badSink(data);
}
\end{lstlisting}
\texttt{<EXAMPLE END>}

Your answer:
The vulnerable example reads input from a file, which could potentially be controlled by an attacker, especially if the file name (FILENAME) or content can be influenced by external input. This input is then used without further validation or sanitization. While the provided code snippet doesn't directly show the execution of a system command with data, it passes data to another function . Given the context, it is implied that data may be used in constructing and executing a system command. The vulnerability stems from concatenating potentially unsafe external input directly to command strings. The code lacks any form of validation or sanitization of the input it reads from the file. This omission allows all characters, including those that could be used maliciously to alter command execution (e.g., ;, \&, |). The snippet doesn't demonstrate the use of safer APIs that treat command parameters as separate entities, which would prevent command injection. Instead, the direct use of input in command execution contexts suggests the absence of such practices. There's no evidence of parameterized commands being used, which would ensure separation of code and data, mitigating injection risks. The provided code doesn't explicitly mention external libraries, but the risk remains if such libraries are involved elsewhere in handling data for command execution without proper precautions.

Q: Check for any vulnerable sections of code in the new example, then determine whether the new example is vulnerable or non-vulnerable based on the instructions.

\texttt{<EXAMPLE START>}
\begin{lstlisting}[basicstyle=\footnotesize\ttfamily, breaklines=true, breakatwhitespace=true, columns=fullflexible, keepspaces=true, frame=single, language=C, xleftmargin=0.05\textwidth, xrightmargin=0.05\textwidth]
static void goodG2B()
{
   wchar_t * data;
   wchar_t dataBuffer[100] = COMMAND_ARG2;
   data = dataBuffer;
   /* FIX: Append a fixed string to data (not user / external input) */
   wcscat(data, L"*.*");
   CWE78_OS_Command_Injection__
    wchar_t_file_w32_execv_
    53b_goodG2BSink(data);
}
\end{lstlisting}
\texttt{<EXAMPLE END>}

Your answer:
The fixed example avoids using external or user input directly in command construction. Instead, it appends a fixed string (\texttt{L"*.*"}) to data, ensuring that the content of data is controlled and predictable. Since the function only appends a predefined string to data, it avoids the risk of command injection that comes from concatenating user-controlled input. The implication is that data can no longer be manipulated to alter command execution maliciously. By not relying on external input for command execution and only appending a safe, known string, the example inherently validates and sanitizes the data used in command contexts. This approach negates the need for explicit sanitization routines for this specific operation. While the code snippet does not explicitly show the execution mechanism, the approach of appending a harmless, static string suggests a safer handling pattern that does not facilitate injection vulnerabilities. The nature of the fix does not directly correlate to parameterized commands but demonstrates a principle of separating data from executable code, which aligns with the intent behind parameterized commands. Similar to the vulnerable function, there's no explicit mention of external libraries. However, the fixed approach does not open opportunities for libraries to execute commands with unsafe input.

Q: Check for any vulnerable sections of code in the new example, then determine whether the new example is vulnerable or non-vulnerable based on the instructions.

\texttt{<NEW EXAMPLE START>}
\begin{lstlisting}[basicstyle=\footnotesize\ttfamily, breaklines=true, breakatwhitespace=true, columns=fullflexible, keepspaces=true, frame=single, language=Python, xleftmargin=0.05\textwidth, xrightmargin=0.05\textwidth]
def _modify_3par_iscsi_host(self, hostname, iscsi_iqn):
    # when using -add, you can not send the persona or domain options
    self.common._cli_run('createhost -iscsi -add %s %s'
                         % (hostname, iscsi_iqn), None)
\end{lstlisting}
\texttt{<NEW EXAMPLE END>}

Your answer:

\end{tcolorbox}

\newpage
\begin{figure*}[t]
  \centering
  \includegraphics[width=\textwidth]{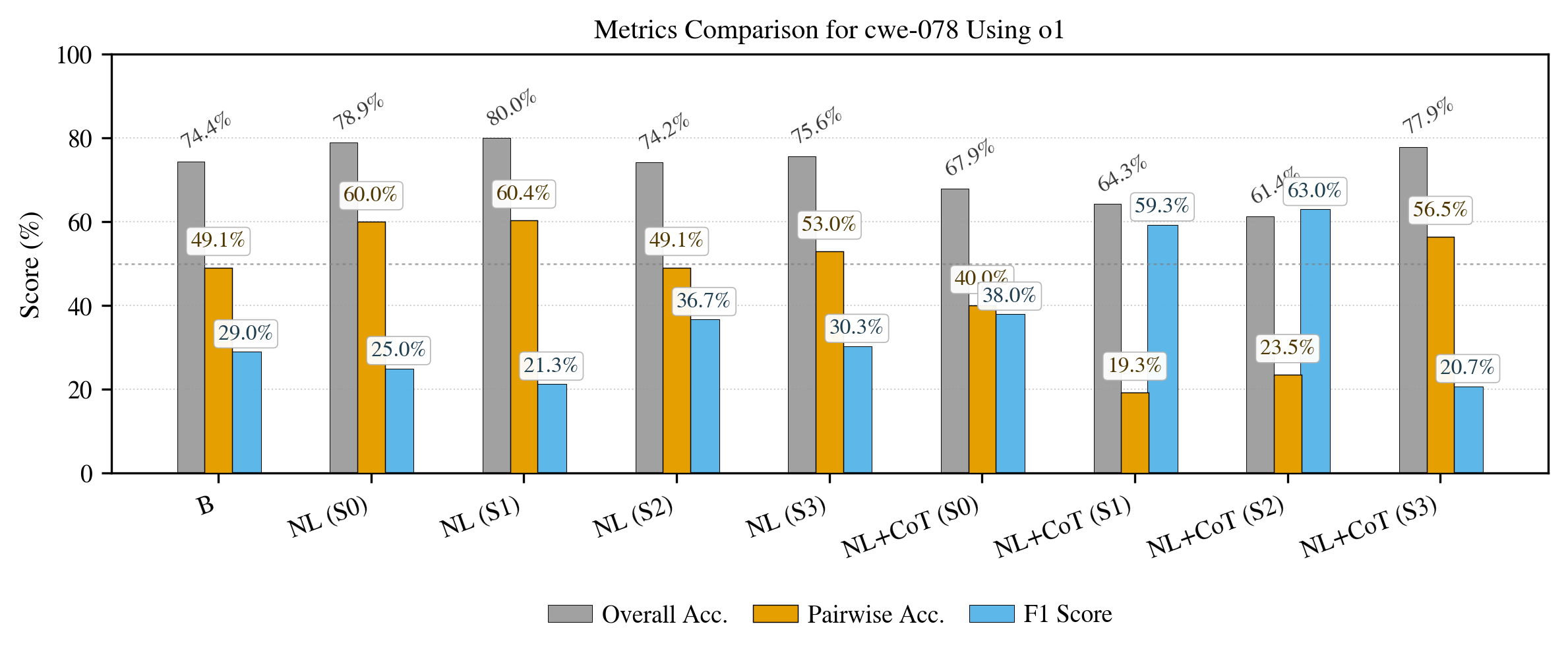}
  \caption{CWE-078 — OpenAI o1}
  \label{fig:cwe078-o1}
\end{figure*}

\begin{figure*}[t]
  \centering
  \includegraphics[width=\textwidth]{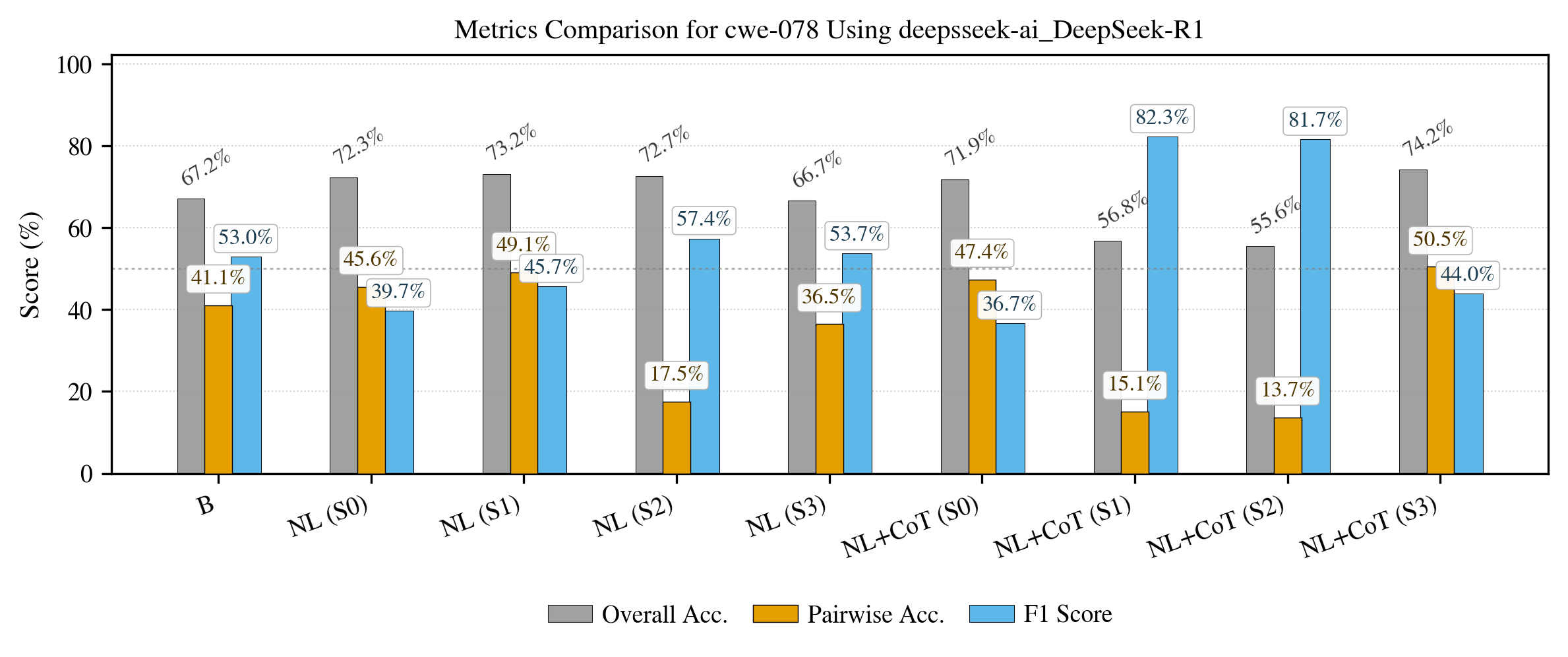}
  \caption{CWE-078 — DeepSeek-R1}
  \label{fig:cwe078-deepseekr1}
\end{figure*}

\begin{figure*}[t]
  \centering
  \includegraphics[width=\textwidth]{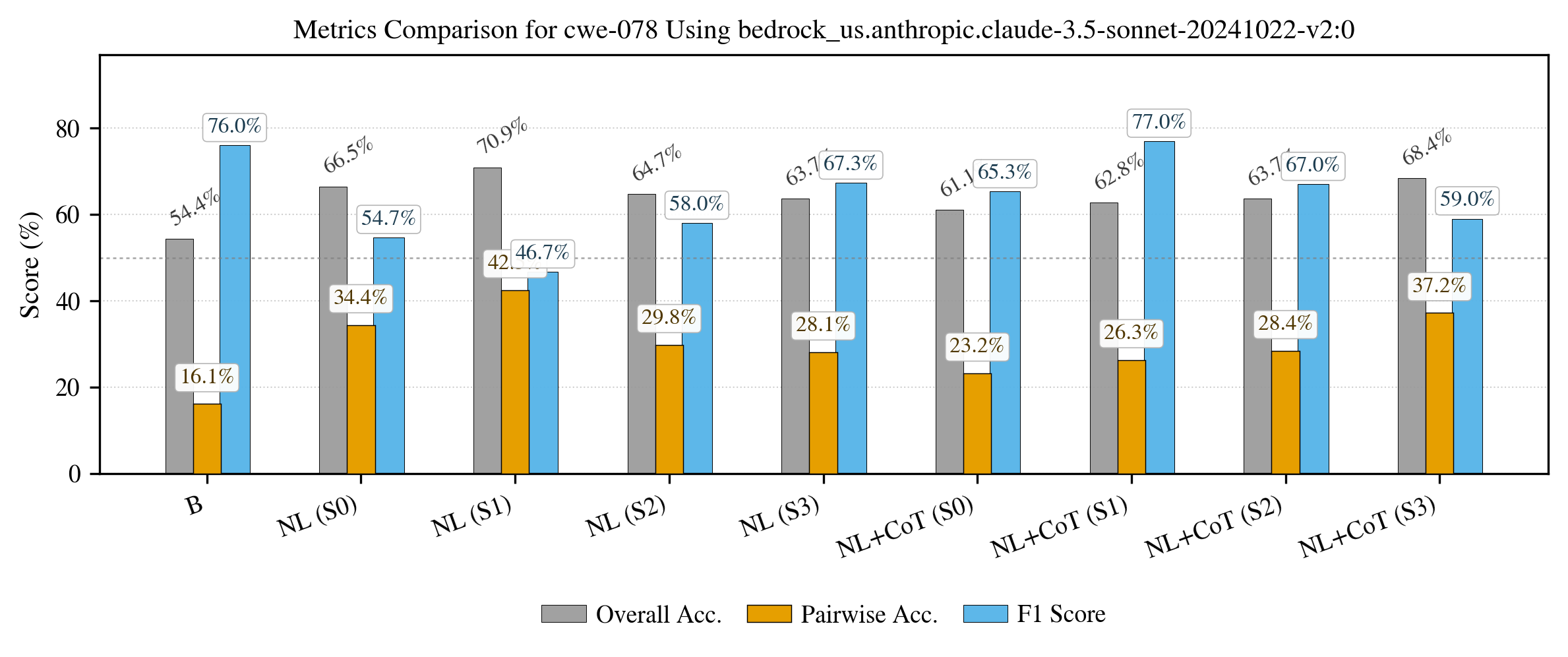}
  \caption{CWE-078 — Claude 3.5 Sonnet v2}
  \label{fig:cwe078-claude35v2}
\end{figure*}

\begin{figure*}[t]
  \centering
  \includegraphics[width=\textwidth]{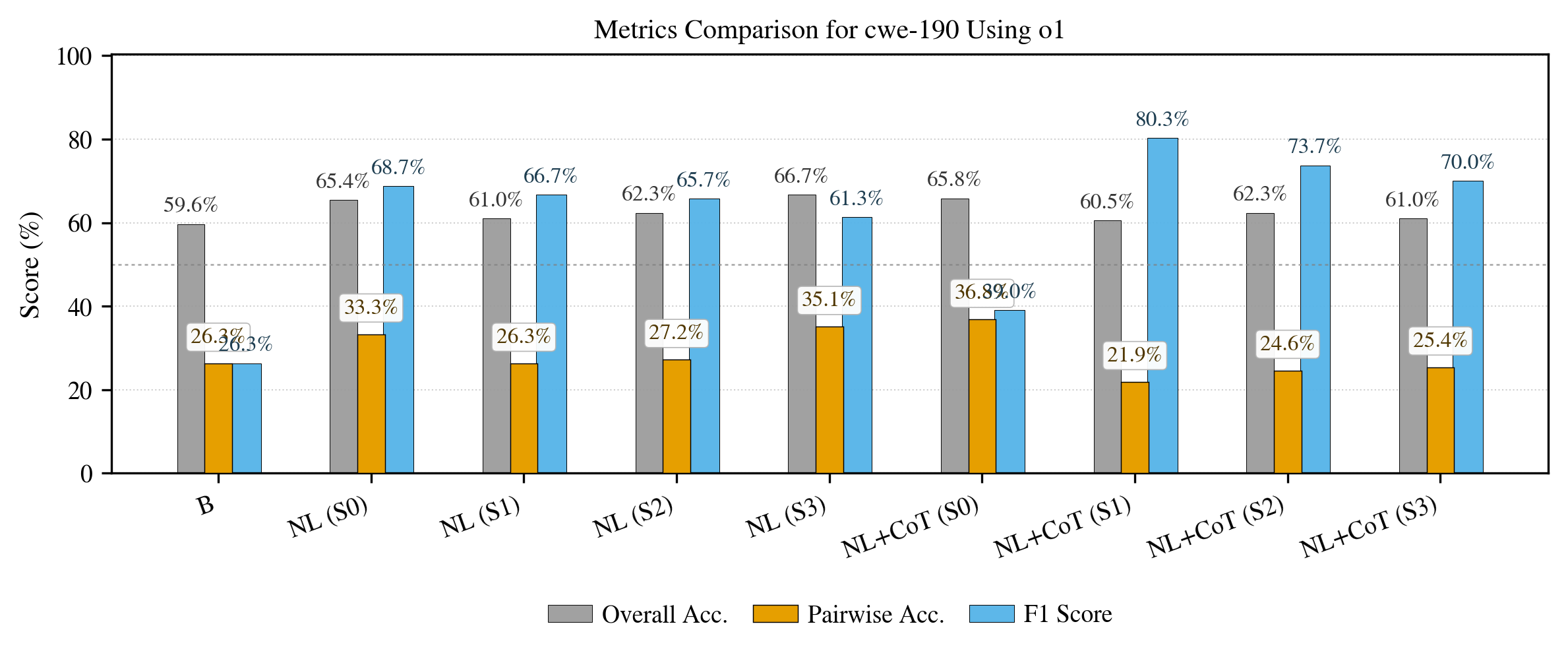}
  \caption{CWE-190 — OpenAI o1}
  \label{fig:cwe190-o1}
\end{figure*}

\begin{figure*}[t]
  \centering
  \includegraphics[width=\textwidth]{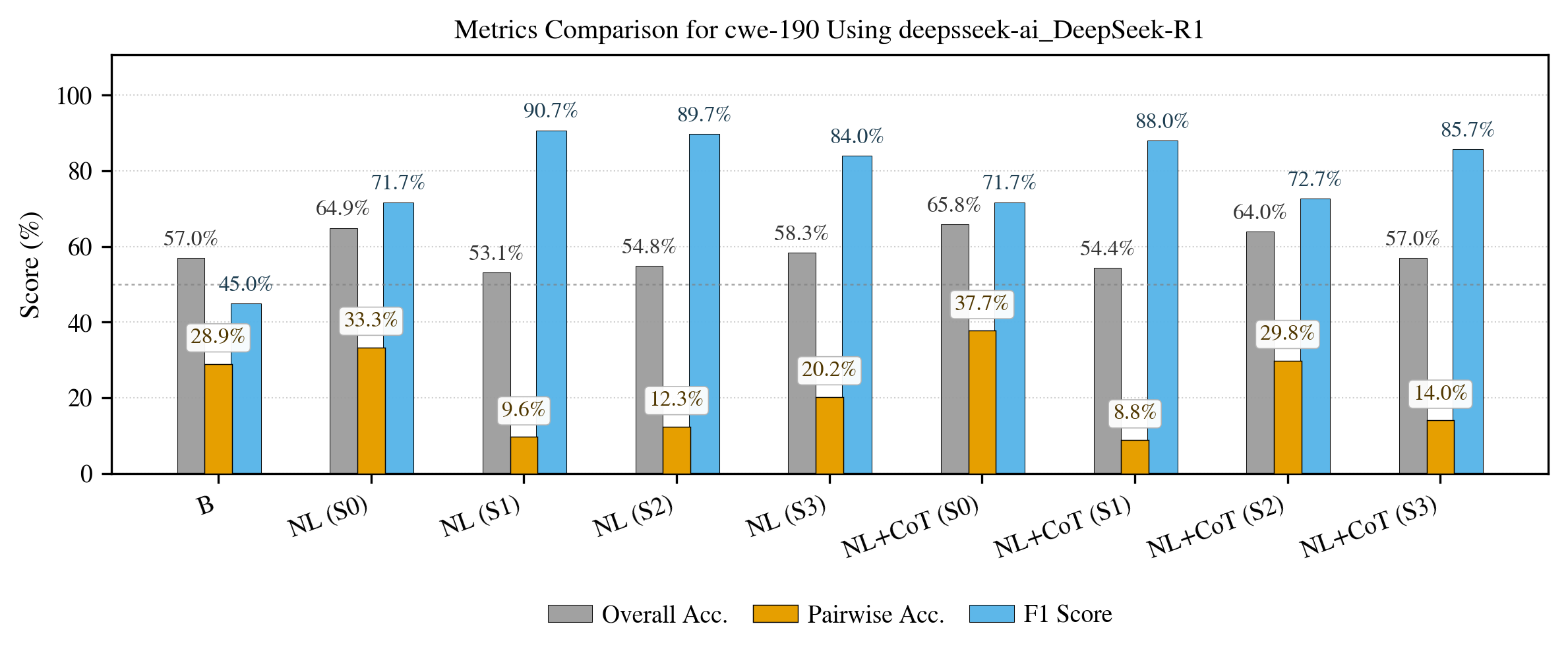}
  \caption{CWE-190 — DeepSeek-R1}
  \label{fig:cwe190-deepseekr1}
\end{figure*}

\begin{figure*}[t]
  \centering
  \includegraphics[width=\textwidth]{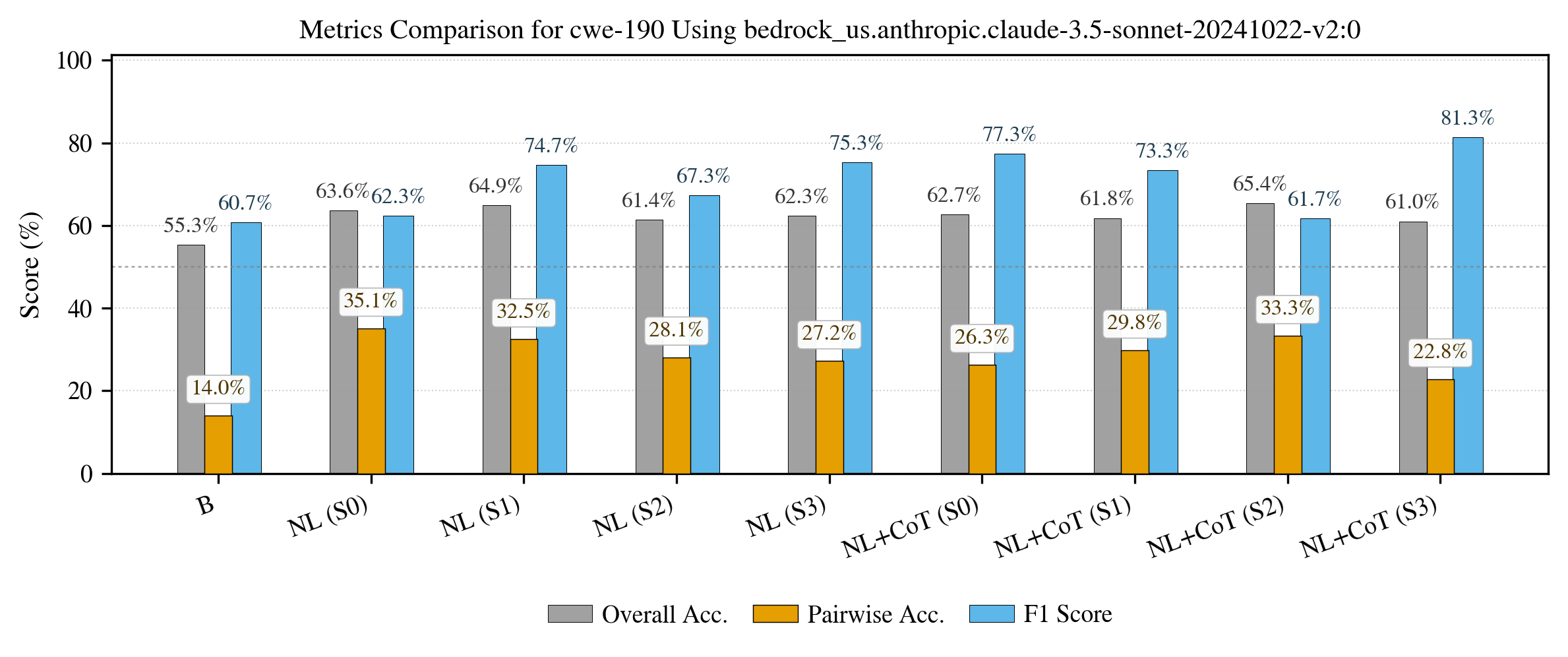}
  \caption{CWE-190 — Claude 3.5 Sonnet v2}
  \label{fig:cwe190-claude35v2}
\end{figure*}

\begin{figure*}[t]
  \centering
  \includegraphics[width=\textwidth]{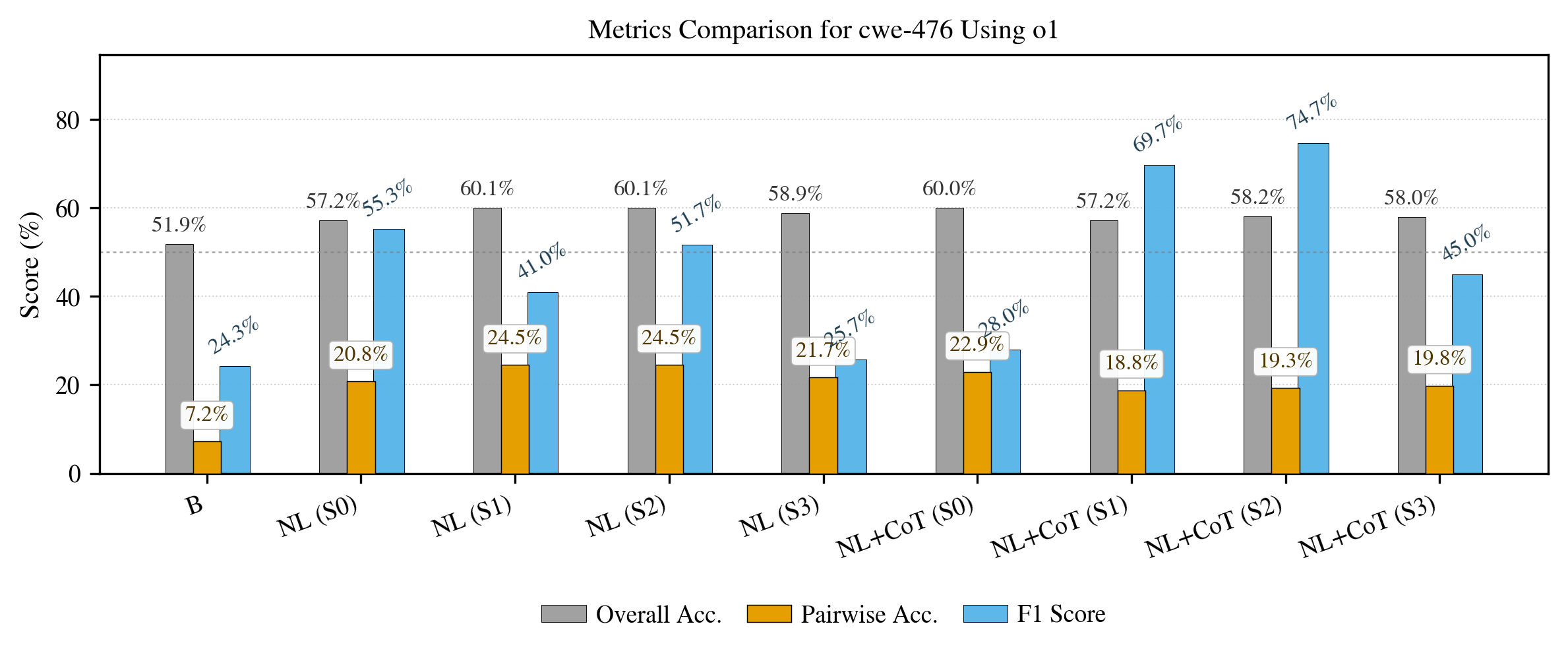}
  \caption{CWE-476 — OpenAI o1}
  \label{fig:cwe476-o1}
\end{figure*}

\begin{figure*}[t]
  \centering
  \includegraphics[width=\textwidth]{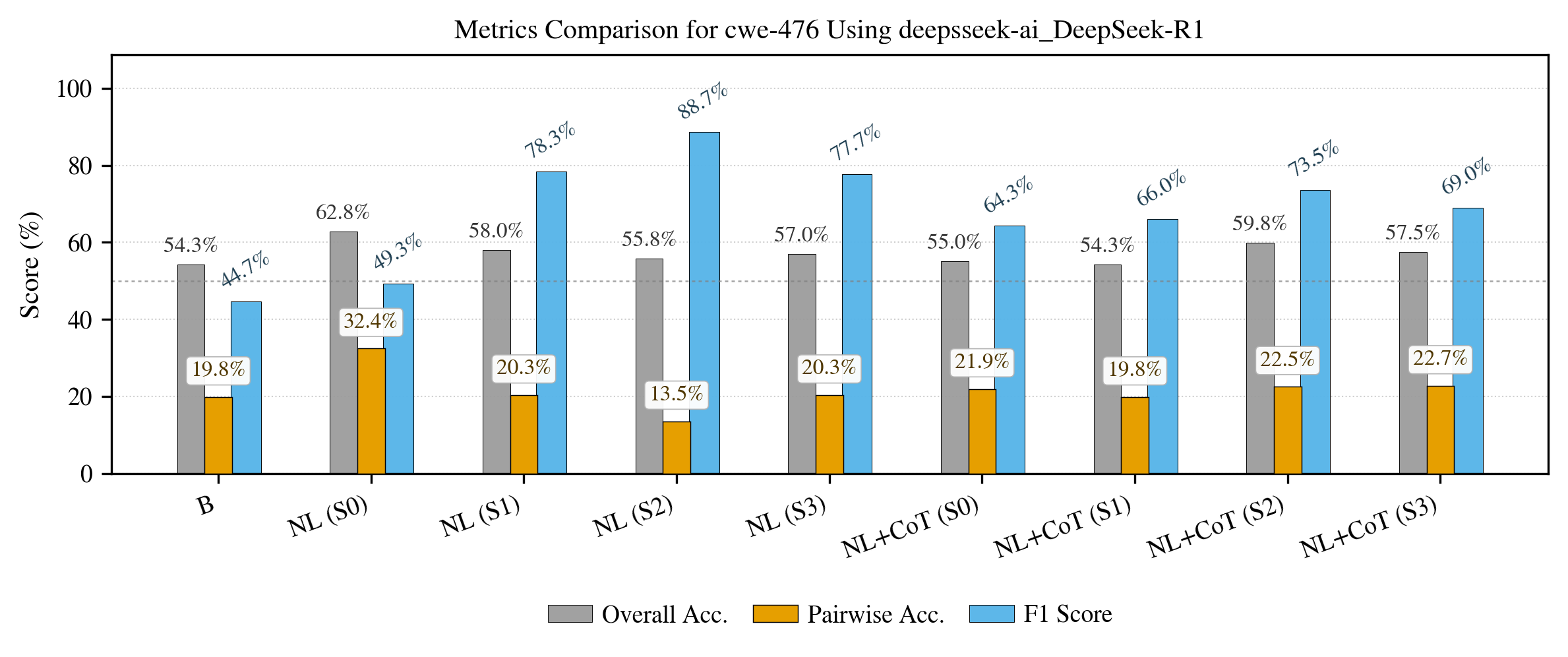}
  \caption{CWE-476 — DeepSeek-R1}
  \label{fig:cwe476-deepseekr1}
\end{figure*}

\begin{figure*}[t]
  \centering
  \includegraphics[width=\textwidth]{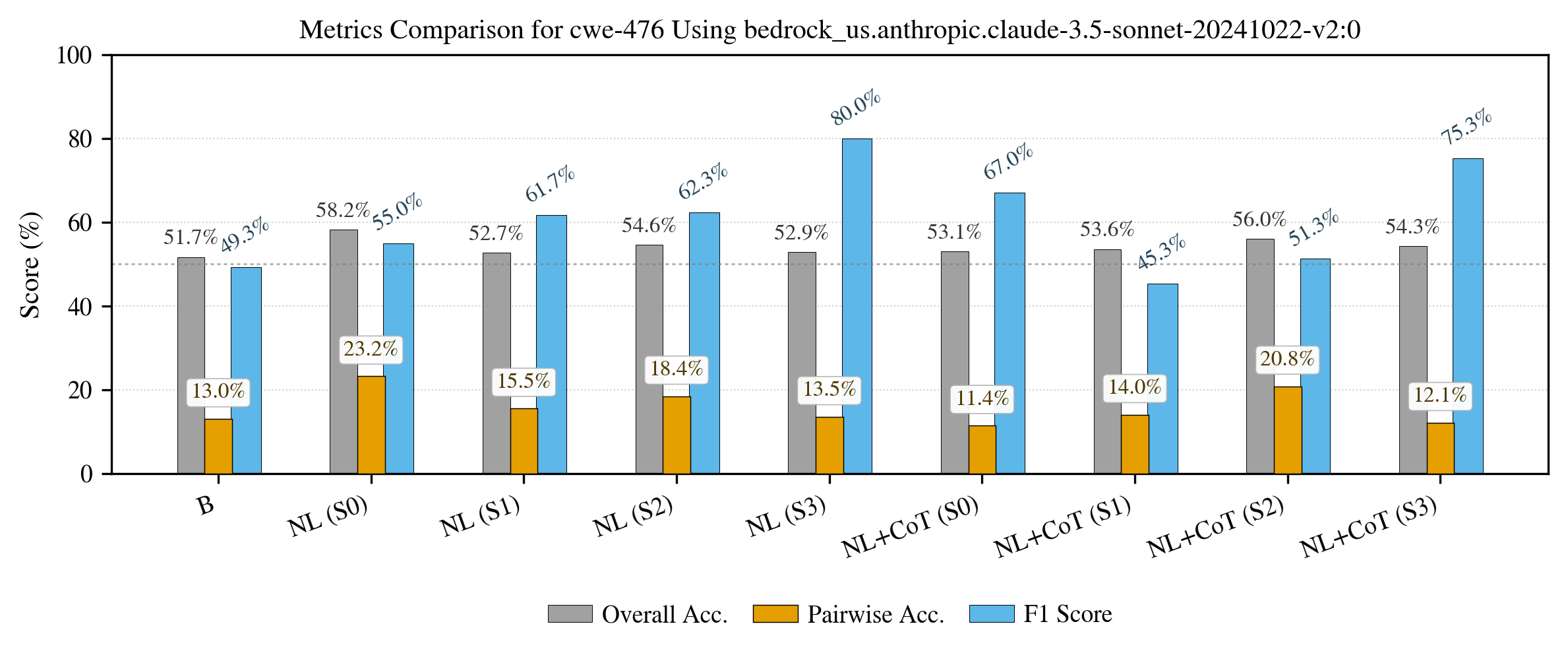}
  \caption{CWE-476 — Claude 3.5 Sonnet v2}
  \label{fig:cwe476-claude35v2}
\end{figure*}

\begin{figure*}[t]
  \centering
  \includegraphics[width=\textwidth]{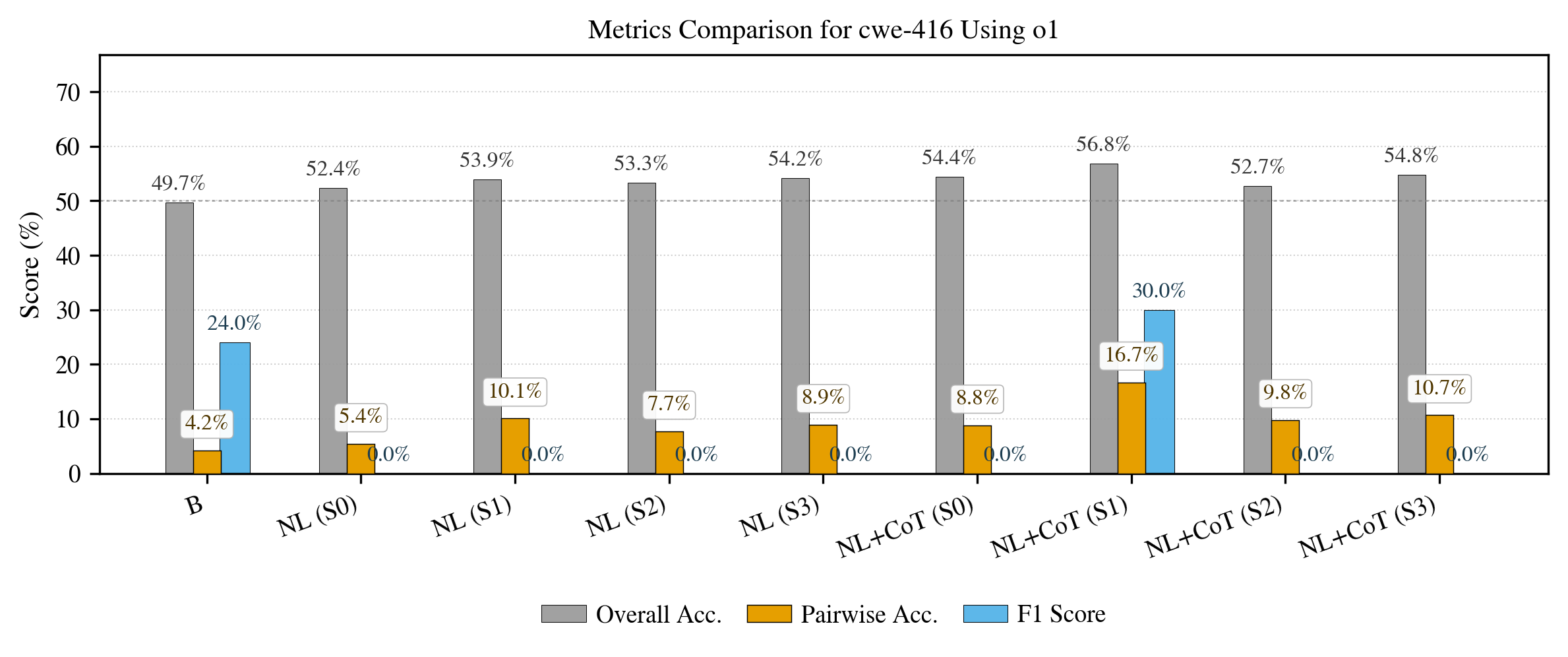}
  \caption{CWE-416 — OpenAI o1}
  \label{fig:cwe416-o1}
\end{figure*}

\begin{figure*}[t]
  \centering
  \includegraphics[width=\textwidth]{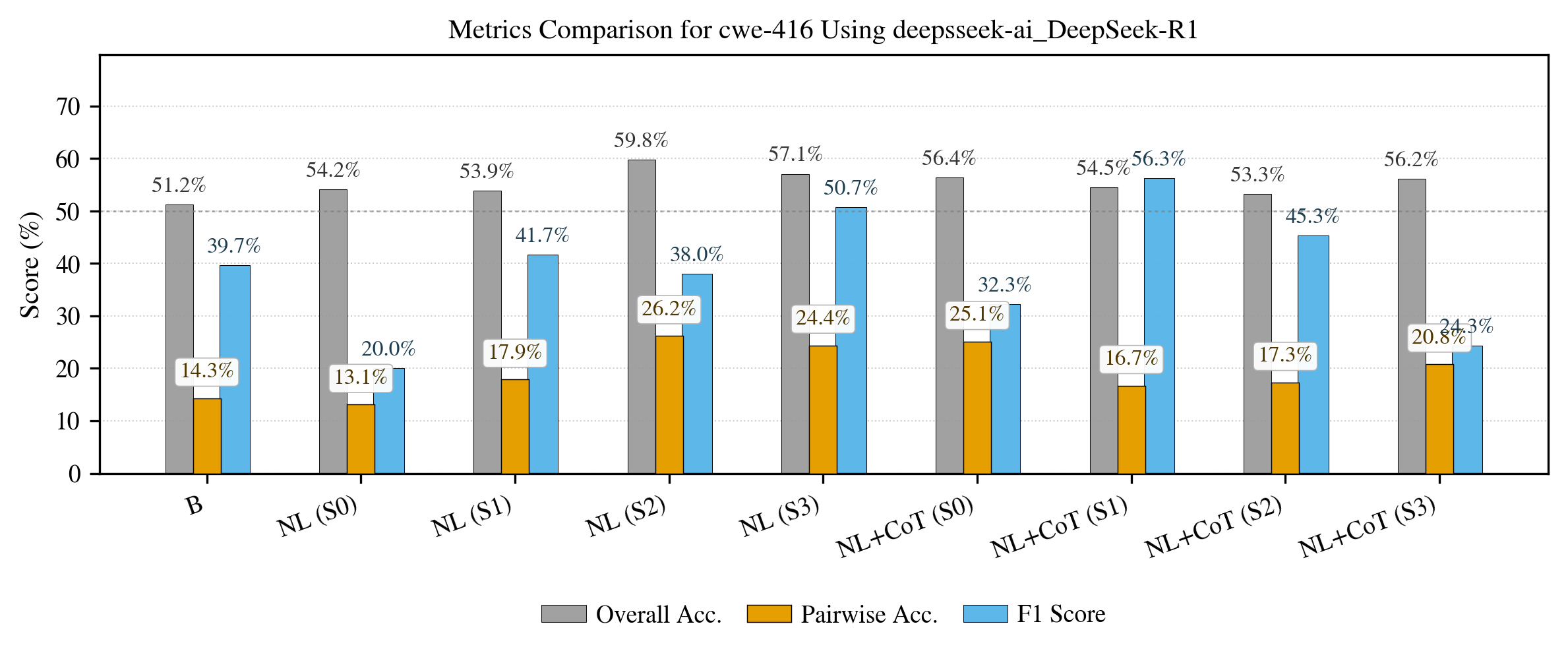}
  \caption{CWE-416 — DeepSeek-R1}
  \label{fig:cwe416-deepseekr1}
\end{figure*}

\begin{figure*}[t]
  \centering
  \includegraphics[width=\textwidth]{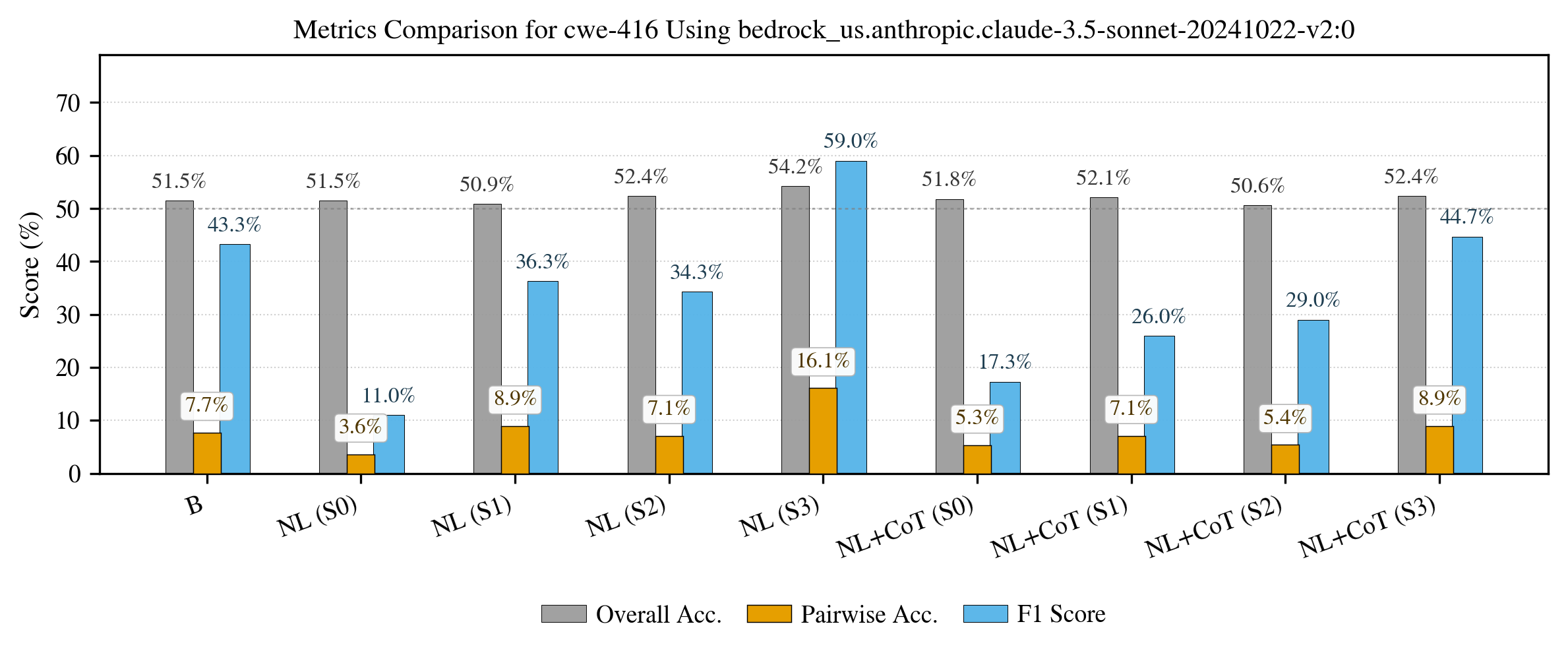}
  \caption{CWE-416 — Claude 3.5 Sonnet v2}
  \label{fig:cwe416-claude35v2}
\end{figure*}

\end{document}